\documentclass[]{aa}  

\usepackage{natbib}
\usepackage{amsmath}
\usepackage{orcidlink}
\usepackage{txfonts}
\usepackage{bm}
\bibpunct{(}{)}{;}{a}{}{,} % to follow the A&A style

\usepackage{color}
\usepackage{graphicx}
\usepackage{txfonts}
\newcommand{\beq}{\begin{equation}}
\newcommand{\eeq}{\end{equation}}
\newcommand{\bea}{\begin{eqnarray}}
\newcommand{\eea}{\end{eqnarray}}

\newcommand{\cm}{\mathcal}
\newcommand{\mb}{\mathbf}
\newcommand{\al}{\langle}
\newcommand{\ar}{\rangle}

\definecolor{azul}{rgb}{0,0,.8}
\definecolor{rojo}{rgb}{1,0,0}
\definecolor{verde}{rgb}{0,.5,0}
\definecolor{violeta}{rgb}{.5,.0,1}
\definecolor{gris}{rgb}{.5,.5,.5}
\definecolor{marron}{rgb}{.4,.1,0}
\definecolor{naranja}{rgb}{1,.5,0}
\definecolor{bordo}{rgb}{.5,0,.2}
\definecolor{rojo}{rgb}{1,0,0}

\begin{document}

\title{Hydrogen photoionization in a magnetized medium:
the rigid-wavefunction approach revisited}

\titlerunning{Hydrogen photoionization opacity in a magnetized medium}

\author{Ren\'e D. Rohrmann\orcidlink{0000-0001-7209-3574}}
 
\institute{
Instituto de Ciencias Astron\'omicas, de la Tierra y del Espacio (ICATE, CONICET-UNSJ), C.C. 467, 5400, San Juan, Argentina
\and
Facultad de Ciencias Exactas, Físicas y Naturales (FCEFN, UNSJ), Av. Ignacio de la Roza 590 (O), J5402DCS San Juan, Argentina
}

\abstract{
Realistic modeling of stellar spectra requires accurate radiative opacity coefficients. Owing to the fragmentary nature of existing data from rigorous quantum-mechanical calculations, photoionization coefficients based on the rigid-wavefunction approximation remain the only practical option for studies of magnetic white dwarfs. Although variants of this approach have been widely used in spectral analyses for decades, a complete and explicit treatment of degeneracy-level breaking has not previously been presented.
In this work, we provide a comprehensive description of this procedure, including explicit expressions for the photoionization probability of individual bound–free transitions as functions of magnetic field strength and radiation polarization. We also evaluate the occupation numbers of bound states in a magnetized gas under ionization equilibrium, enabling the calculation of absolute photoionization opacities. Because high-lying atomic states are strongly perturbed by the magnetic field and ultimately dissolved, substantial modifications of the monochromatic absorption are found even for fields below 10 MG—a regime where fully rigorous quantum calculations are numerically demanding and have not yet been applied. Over a wide range of magnetic field strengths, pronounced dichroic features appear in the hydrogen continuum absorption.
} 

\keywords{atomic processes --- magnetic fields --- opacity --- stars: atmospheres --- white dwarfs}

\maketitle
%
%________________________________________________________________

%%%%%%%%%%%%%%%%%%%%%%%%%%%%%%%%%%%%%%%%%%%%%%%%%%%%%%%%%%%%%%%%%
\section{Introduction}\label{s:intro}

Magnetic fields at the surfaces of magnetic white dwarf stars (MWDs) have been reported to span values of approximately $10^{-2}$--$10^3$~ (1~MG~$=10^6$~G). Such fields can significantly affect the physical properties of the gas, the transport of energy in the stellar atmosphere -- usually composed of pure hydrogen -- and the distribution of radiation emitted by the star. A number of studies have investigated photoabsorption processes by atoms in the presence of magnetic fields of this strength.

An approximate treatment of hydrogen photoionization in magnetized photospheres was first formulated by \citet{lamb1972, lamb1974} for weak fields, assuming linear Zeeman theory. During the 1980s, detailed theoretical evaluations of bound–free transitions were carried out for very high magnetic field strengths (upper $10^3$~MG), slightly exceeding the regime of presently known MWDs \citep{greene1983, battacharya1985}. Subsequently, a limited number of studies based on fully quantum-mechanical methods have been published, reporting hydrogen photoionization cross sections for magnetic fields above 10~MG \citep{wang1991} and 20~MG \citep{alijah1990, delande1991, merani1995, meinhardt1999, zhao2007, zhao2021}.\footnote{Other studies have focused on photoabsorption in the intense magnetic fields at the surfaces of neutron stars (e.g., \citealt{gnedin1974, schmidt1981, potekhin1997}). These works typically employ specialized wavefunction bases to represent the energy eigenfunctions, which are not well suited to the magnetic field strengths characteristic of} MWDs. Rigorous calculations at lower magnetic field strengths remain lacking, in part because the associated technical difficulties are greater than in the high-field regime \citep{mota2007}.

%=============================================
\begin{figure}
\includegraphics[width=.5\textwidth]{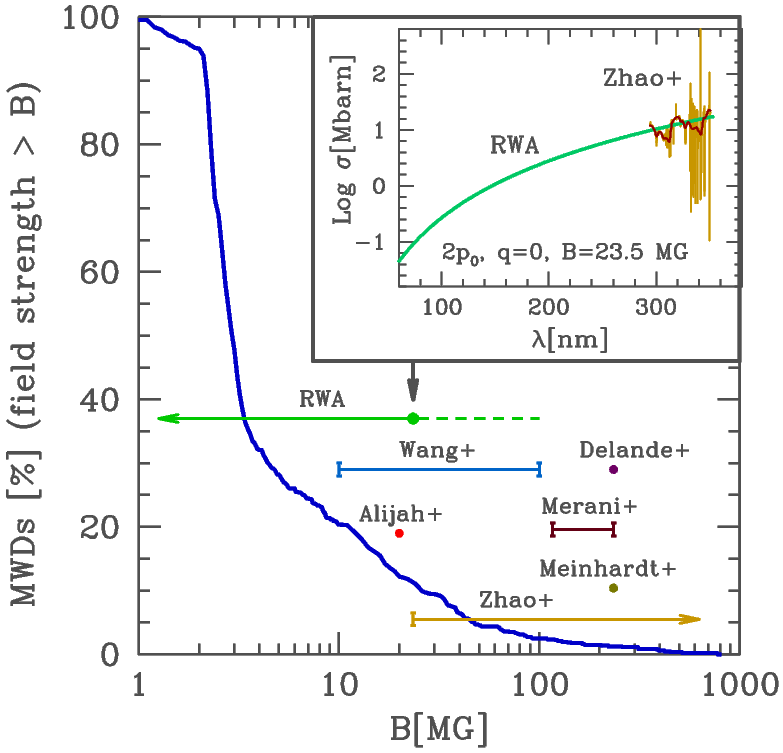}
\caption{Cumulative distribution (in percentage) of known MWDs with mean field strengths greater than $B$ (thick blue line). Field strength ranges analized in photoionization studies are indicated.
The insert figure shows a cross-section evaluated by a fully quantum-mechanic method \citep{zhao2007} and the results obtained by the RWA approach (see text). The dark line represents a Gaussian convolution of the Zhao \& Stancil evaluation with a FWHM of 3.5~nm~ ($\approx 10^{-3}$~Ry).}
\label{f:NB}
\end{figure}
%=============================================
Fig.~\ref{f:NB} shows the fraction of known MWDs with field strengths above a given value $B$ in the range 1--$10^3$~MG, based on a sample of 804 objects -- the largest compilation of MWDs with measured magnetic fields to date \citep{amorim2023}. Fully quantum-mechanical evaluations of bound–free absorption have focused on conditions characteristic of highly magnetized objects, which represent only a small fraction (10\%--20\%) of presently known MWDs.
The situation is even more critical given that the number of transitions studied using such methods typically includes, at best, only a few dozen over a limited photon-energy range. This is insufficient to quantitatively account for the opacities required in spectral modeling. Magnetic fields break the degeneracy of electronic states and introduce dichroism, that is, a dependence of the cross section on photon polarization. Consequently, the calculation of synthetic spectra using model atmospheres requires cross-section data for several hundred to thousands of transitions. To date, no MWD spectrum can be fully analyzed using the sparse data available from rigorous quantum-mechanical techniques. In addition, cross sections are required over a wide photon-frequency range and on a fine grid of magnetic field strengths, since a uniform field is not expected across the stellar disk.

The approach proposed by \citet{lamb1972,lamb1974} therefore remains the only method capable of providing photoionization cross sections for MWD model atmospheres. Their theory relies on a key approximation: the bound- and free-electron wavefunctions are assumed to be unchanged by the magnetic field in the small region of overlap, so that the matrix elements governing radiative transitions are unaffected. This approach is known as the rigid-wavefunction approximation (RWA). The photoionization cross section also depends on the transition energy, which is modified by the magnetic field through the shifting and splitting of the initial and final electron eigenenergies. In their original formulation, Lamb \& Sutherland considered only bound-level shifts described by linear Zeeman theory.
Later, \citet{jordan1989} demonstrated that the photoabsorption spectrum of magnetized hydrogen exhibits multiple absorption edges associated with different initial states and continuum thresholds. As improved atomic data became available, more accurate bound-state energies were incorporated into RWA-based calculations. However, the complete procedure was never explicitly described, despite its use in several numerical codes developed to model MWD atmospheres \citep{martin1986, jordan1992, putney1995, wickramasinghe2000, euchner2002, vera2024}.

Although RWA is strictly valid only in the zero-field limit, it has proven to be a reliable empirical method for providing average photoionization cross sections at moderate field strengths \citep{rohrmann2025}. The inset in Fig.~\ref{f:NB} compares RWA predictions with the detailed calculations of \citet{zhao2007} for a specific transition at the lowest magnetic field strength considered in that work. The primary difference between the two results arises from the rich resonance structure showed by the full quantum treatment. In MWD atmospheres, however, these resonances are expected to be smeared into a smooth continuum by field variations across the stellar surface \citep{wickramasinghe1995}. As shown in Fig.~\ref{f:NB}, RWA reproduces the mean behavior of the full theory for the transition from the $2p_0$ 
state induced by absorption of a linearly polarized photon ($q=0$) at $B=23.5$~MG, a field strength well beyond the linear Zeeman regime.

The present paper aims to provide a comprehensive evaluation of the photoionization opacity of hydrogen atoms using RWA. This objective is motivated by two main considerations. First, RWA is currently the only available method for obtaining a complete set of photoionization cross sections suitable for modeling MWD spectra over a wide range of magnetic field strengths. Second, its full implementation -- accounting for weighted sums of partial cross sections that depend on light polarization and on photoionization branching fractions (i.e., the relative probabilities of ionization into specific final states) -- has never been explicitly detailed. In addition, we determine the ionization equilibrium of magnetized hydrogen gas and compute the occupation numbers of bound states required to derive absolute photoionization opacities.

The following sections address the key elements involved in calculating absolute opacity using RWA photoionization within the electric-dipole approximation. Section~\ref{s:partial} presents the decomposition of the cross section into multiple components, whose statistical weights are concisely expressed using the Wigner-Eckart theorem. In Sect.~\ref{s:branching}, branching fractions are obtained through analytic continuation of bound–bound oscillator strengths. Section~\ref{s:rigid} describes the evaluation of electron eigenenergies, continuum-threshold shifts, and occupation numbers of initial states for a partially ionized hydrogen gas as functions of magnetic field strength. The results are discussed in Sect.~\ref{s:results}, and conclusions are drawn in Sect.~\ref{s:concl}.

%%%%%%%%%%%%%%%%%%%%%%%%%%%%%%%%%%%%%%%%%%%%%%%%%%%%%%%%%%%%%%%%%
\section{Partial photoionization cross-sections}\label{s:partial}

The absorption cross section in the electric-dipole approximation for a one-electron system is proportional to the photon energy $\cm{E}$ and to the square of the modulus of the transition matrix element,
\beq \label{basic}
\sigma^q_{nlm,kl'm'} = \text{cte} \; \cm{E}
        |\al nlm |\mb{r}.\mb{e}_q| kl'm'\ar|^2.
\eeq
Here $|nlm\ar$ and $|kl' m'\ar$  denote the initial (bound) and final (ionized) atomic states, characterized by the principal ($n$), quasi-principal ($k$), orbital ($l,l'$) and magnetic ($m,m'$) quantum numbers. The vector $\mb{r}$ is the electron position operator, and $\mb{e}_q$ is the unit polarization vector of the photon,
\beq\label{d.e}
\mb{e}_0=\mb{e}_z
,\hskip.2in
\mb{e}_{\pm1}=\mp\frac{1}{\sqrt{2}}\left(\mb{e}_x \pm i \mb{e}_y\right),
\eeq
where $\hbar q$ is the $z$ component of the angular momentum carried by the photon, with $q=0,\pm \kappa$ ($\kappa=1$). As expressed by Eq.~(\ref{d.e}), $q=0$  corresponds to linearly polarized radiation (parallel to the $z$-axis), while $q=\pm 1$ corresponds to right- ($+$) or left-handed ($-$) circularly polarized radiation in the $xy$ plane. These transitions are conventionally referred to as $\pi$, $\sigma^+$ and $\sigma^-$, respectively.

For field-free atoms, the matrix element in Eq.~(\ref{basic}) can be evaluated using exact nonrelativistic wavefunctions,  $\al \mb{r}|nlm\ar = R_{nl}(r)Y_{lm}(\theta,\phi)$ and $\al \mb{r}|kl'm'\ar = R_{kl'}(r)Y_{l'm'}(\theta,\phi)$, which depend on the polar coordinates ($r,\theta,\phi$) \citep{stobbe1930, bethe1957}. Alternatively, it may be expressed using the Wigner-Eckart theorem,
\beq
\left\al nlm\left|\mb{r}.\mb{e}_q\right|kl'm'\right\ar =(-1)^{l-m} 
\left(\begin{array}{ccc} l& \kappa & l' \\ -m& q & m' \end{array}\right)
   \left\al nl\left\|r\right\|kl'\right\ar ,
\eeq
where the integration over the angular degrees of freedom has already been carried out. The cross section,
\beq \label{1a}
\sigma^q_{nlm,kl'm'} = \text{cte}\; \cm{E} 
      \cm{W}^{\;l,\;\kappa,\; l'}_{m,\, q,\,m'}
        |\al nl\|r\| kl'\ar|^2,
\eeq
then depends on the reduced matrix element $|\al nl\|r\| kl'\ar|$  and on the square of the Wigner 3$j$ coefficient,
\beq
\cm{W}^{\;l,\;\kappa,\; l'}_{m,\, q,\,m'} \equiv
\left(\begin{array}{ccc} l& \kappa & l' \\ m& q &-m' \end{array}\right)^2.
\eeq
Values of interest for the current transitions are the following 
\bea
\cm{W}^{\;l,\;1,\; l+1}_{m,\,-1,\,m-1}  &=& 
    \frac{(l-m+2)(l-m+1)}{2(l+1)(2l+1)(2l+3)}, \cr  && \,^{\,}\cr
\cm{W}^{\;l,\;1,\; l+1}_{m,\,0,\,m} &=&
    \frac{(l+1)^2-m^2}{(l+1)(2l+1)(2l+3)}, \cr && \,^{\,}\cr
\cm{W}^{\;l,\;1,\; l+1}_{m,\,1,\,m+1} &=&
    \frac{(l+m+2)(l+m+1)}{2(l+1)(2l+1)(2l+3)}, \cr && \,^{\,}\cr
\cm{W}^{\;l,\;1,\; l-1}_{m,\,-1,\,m-1} &=&
    \frac{(l+m)(l+m-1)}{2l(2l+1)(2l-1)},  \cr && \,^{\,}\cr
\cm{W}^{\;l,\;1,\; l-1}_{m,\,0,\,m} &=&
    \frac{l^2-m^2}{l(2l+1)(2l-1)}, \cr && \,^{\,}\cr
\cm{W}^{\;l,\;1,\; l-1}_{m,\,1,\,m+1} &=&
    \frac{(l-m)(l-m-1)}{2l(2l+1)(2l-1)}.
\eea
The selection rules for electric-dipole transitions, obtained either by direct evaluation of Eq.~(\ref{basic}) or inferred from the properties of the Wigner 3$j$ coefficients, are
\beq\label{selection}
l'-l = \pm \kappa= \pm 1,\hskip.3in m'-m = q =0,\pm 1.
\eeq

The cross section for nonoriented atoms is obtained by averaging over the degeneracy of the initial-state orientations (i.e., over $m$ values) and summing over the degeneracy of the final states ($-l'\le m'\le l'$),
\beq \label{1b}
\sigma_{nl,kl'}=\frac{1}{2l+1}\sum_{m m'} \sigma^q_{nlm,kl'm'}
=\text{cte} \frac{\cm{E}}{3(2l+1)} |\al nl\|\mb{r}\| kl'\ar|^2,
\eeq
where the final equality follows from the summation property
\beq \label{1c}
\sum_{m,m'} \cm{W}^{\;l,\;\kappa,\; l'}_{m,\, q,\,m'}=\frac{1}{2\kappa+1}=\frac13.
\eeq
By combining Eqs. (\ref{1a}) and (\ref{1b}), one obtains
\beq \label{sigwig}
\sigma^q_{nlm,kl'm'}=3(2l+1)
\cm{W}^{\;l,\;\kappa,\; l'}_{m,\, q,\,m'} \sigma_{nl,kl'}.
\eeq
We define $\xi_{nl,kl'}$ as the branching ratio for photoionization transitions $nl\rightarrow kl'$, that is, the fraction of photoionizations from a given bound sublevel ($nl$) into the specific continuum channel ($kl'$), relative to all photoionizations originating from the level $n$. In terms of cross sections,
\beq\label{branching}
\sigma_{nl,kl'}=\xi_{nl,kl'}\sigma_{n,k},
\eeq
with
\beq\label{sigmank}
\sigma_{n,k}=\frac{1}{n^2}\sum_{ll'}(2l+1)\sigma_{nl,kl'}.
\eeq
Equations~(\ref{sigwig}) and (\ref{branching}) allow the total cross section from an initial state $nlm$ to be expressed for all allowed electric-dipole transitions ($l'=l\pm1$) and for a specific photon polarization $q$ ($m'=m+q$) as
\beq \label{e.main}
\sigma^q_{nlm,k} = 3(2l+1) 
 \left[\cm{W}^{\;l,\;\kappa,\; l+1}_{m,\, q,\,m+q} \xi_{nl,k(l+1)}
     + \cm{W}^{\;l,\;\kappa,\; l-1}_{m,\, q,\,m+q} \xi_{nl,k(l-1)}\right] 
  \sigma_{n,k},
\eeq
which follows from applying the selection rules given in Eq.~(\ref{selection}) and summing over all accessible sublevels in the continuum, 
\beq\label{sigmank}
\sigma_{nlm,k}^q = \sum_{l'm'} \sigma_{nlm,kl'm'}^q.
\eeq
%

%=============================================
\begin{table}
\caption{Transition weights  $A^q_{lm}$ [Eq. (\ref{Aqnlm})]. Values for negative $m$ result from the symmetry relations given in Eq. (\ref{ABsymm}).
\label{T.1}}
\setlength{\tabcolsep}{3pt}
\begin{small}
\begin{tabular}{cccccccc}\hline\hline
 $q$  & $m=0$ &  $m=1$ & $m=2$ & $m=3$ & $m=4$ & $m=5$ & $m=6$ \\
\hline
%     &       &        &       &       &       &       &       \\
      &       &        &       &($l=0$)&       &       &       \\
\hline 
$-1$  & 1 &   &   &   &   &   &   \\
$\;0$ & 1 &   &   &   &   &   &   \\
$+1$  & 1 &   &   &   &   &   &   \\
\hline
      &       &        &       &($l=1$)&       &       &       \\
\hline 
$-1$  &  9/10 &  3/10  &   &   &   &   &   \\
$\;0$ &  6/5  &  9/10  &   &   &   &   &   \\
$+1$  &  9/10 &  9/5   &   &   &   &   &   \\
\hline 
      &       &        &       &($l=2$)&       &       &       \\
\hline 
$-1$  &  6/7  &  3/7   &  1/7  &   &   &   &   \\
$\;0$ &  9/7  &  8/7   &  5/7  &   &   &   &   \\
$+1$  &  6/7  & 10/7   & 15/7  &   &   &   &   \\
\hline 
      &       &        &       &($l=3$)&       &       &       \\
\hline 
$-1$  &  5/6  &  1/2   &  1/4  &  1/12 &       &       &       \\
$\;0$ &  4/3  &  5/4   &   1   &  7/12 &       &       &       \\
$+1$  &  5/6  &  5/4   &  7/4  &  7/3  &       &       &       \\
\hline 
      &       &        &       &($l=4$)&       &       &       \\
\hline 
$-1$  &  9/11 &  6/11  & 18/55 &  9/55 &  3/55 &       &       \\
$\;0$ & 15/11 & 72/55  & 63/55 & 48/55 & 27/55 &       &       \\
$+1$  &  9/11 & 63/55  & 84/55 &108/55 & 27/11 &       &       \\
\hline 
      &       &        &       &($l=5$)&       &       &       \\
\hline 
$-1$  & 21/26 & 15/26  &  5/13 &  3/13 &  3/26 &  1/26 &       \\
$\;0$ & 18/13 & 35/26  & 16/13 & 27/26 & 10/13 & 11/26 &       \\
$+1$  & 21/26 & 14/13  & 18/13 & 45/26 & 55/26 & 33/13 &       \\
\hline 
      &       &        &       &($l=6$)&       &       &       \\
\hline 
$-1$  &  4/5  &  3/5   &  3/7  &  2/7  &  6/35 &  3/35 &  1/35 \\
$\;0$ &  7/5  & 48/35  &  9/7  &  8/7  & 33/35 & 24/35 & 13/35 \\
$+1$  &  4/5  & 36/35  &  9/7  & 11/7  & 66/35 & 78/35 & 13/5  \\
\hline\hline
\end{tabular}
\end{small}
%\tablefoot{...}
\end{table}
%=============================================

%=============================================
\begin{table}
\caption{Transition weights $B^q_{lm}$ [Eq. (\ref{Bqnlm})]. Values for negative $m$ result from the symmetry relation given in Eq. (\ref{ABsymm}). Note that $B^q_{00}\equiv 0$ $\forall q$.
\label{T.2}}
\setlength{\tabcolsep}{3pt}
\begin{small}
\begin{tabular}{cccccccc}\hline\hline
 $q$  & $m=0$ &  $m=1$ & $m=2$ & $m=3$ & $m=4$ & $m=5$ & $m=6$ \\
\hline
%     &       &        &       &       &       &       &       \\
      &       &        &       &($l=1$)&       &       &       \\
\hline 
$-1$  &   0   &   3    &      &      &      &      &      \\
$\;0$ &   3   &   0    &      &      &      &      &      \\
$+1$  &   0   &   0    &      &      &      &      &      \\
\hline 
      &       &        &       &($l=2$)&       &       &       \\
\hline 
$-1$  &  1/2  &  3/2   &   3   &      &      &      &      \\
$\;0$ &   2   &  3/2   &   0   &      &      &      &      \\
$+1$  &  1/2  &   0    &   0   &      &      &      &      \\
\hline 
      &       &        &       &($l=3$)&       &       &       \\
\hline 
$-1$  &  3/5  &  6/5   &   2   &   3   &      &      &      \\
$\;0$ &  9/5  &  8/5   &   1   &   0   &      &      &      \\
$+1$  &  3/5  &  1/5   &   0   &   0   &      &      &      \\
\hline 
      &       &        &       &($l=4$)&       &       &       \\
\hline 
$-1$  &  9/14 & 15/14  & 45/28 &  9/4  &   3   &      &      \\
$\;0$ & 12/7  & 45/28  &  9/7  &  3/4  &   0   &      &      \\
$+1$  &  9/14 &  9/28  &  3/28 &   0   &   0   &      &      \\
\hline 
      &       &        &       &($l=5$)&       &       &       \\
\hline 
$-1$  &  2/3  &   1    &  7/5  & 28/15 & 12/5  &   3   &      \\
$\;0$ &  5/3  &  8/5   &  7/5  & 16/15 &  3/5  &   0   &      \\
$+1$  &  2/3  &  2/5   &  1/5  &  1/15 &   0   &   0   &      \\
\hline 
      &       &        &       &($l=6$)&       &       &       \\
\hline 
$-1$  & 15/22 & 21/22  & 14/11 & 18/11 & 45/22 &  5/2  &   3   \\
$\;0$ & 18/11 & 35/22  & 16/11 & 27/22 & 10/11 &  1/2  &   0   \\
$+1$  & 15/22 &  5/11  &  3/11 &  3/22 &  1/22 &   0   &   0   \\
\hline\hline
\end{tabular}
\end{small}
%\tablefoot{...}
\end{table}
%=============================================

%%%%%%%%%%%%%%%%%%%%%%%%%%%%%%%%%%%%%%%%%%%%%%%%%%%%%%%%%%%%%%%%%
\subsection{Rules on geometrical and physical factors}\label{s:factors}

Equation (\ref{e.main}) can be rewritten as
\beq \label{e.main2}
\sigma^q_{nlm,k} =
 \left[A^q_{lm} \xi_{nl,k(l+1)} + B^q_{lm}\xi_{nl,k(l-1)}\right] \sigma_{n,k},
\eeq
with
\beq \label{Aqnlm}
 A^q_{lm} \equiv 3(2l+1) \cm{W}^{\;l,\;\kappa,\; l+1}_{m,\, q,\,m+q},
\eeq
\beq \label{Bqnlm}
 B^q_{lm} \equiv 3(2l+1) \cm{W}^{\;l,\;\kappa,\; l-1}_{m,\, q,\,m+q}.
\eeq
Equation~(\ref{e.main2}) summarizes the separation introduced by the Wigner–Eckart theorem into a geometric contribution, contained in the coefficients $A^q_{lm}$ and $B^q_{lm}$, and a physical contribution, expressed by the branching ratios $\xi_{nl,kl'}$ and the total cross section $\sigma_{n,k}$. Because of the symmetry property
\beq %\label{Wqn}
 \cm{W}^{\;l,\;\kappa,\; l'}_{-m,\,-q,-\,m'} =
  \cm{W}^{\;l,\;\kappa,\; l'}_{m,\, q,\,m'},
\eeq
the coefficients $A$ and $B$ satisfy the following relations under simultaneous sign changes of the indices $q$ and $m$:
\beq \label{ABsymm}
 A^{-q}_{l,-m} = A^{+q}_{l,+m},\hskip.3in
 B^{-q}_{l,-m} = B^{+q}_{l,+m}.
\eeq
From properties of the Wigner 3$j$ symbols it also follows that
\beq 
\sum_q A^q_{lm} =\sum_q B^q_{lm} =2\kappa+1=3,\hskip.3in \forall (l,m),
\eeq
and
\beq 
\sum_m A^q_{lm} =\sum_m B^q_{lm} =2l+1,\hskip.3in \forall (l,q).
\eeq
Tables~\ref{T.1} and \ref{T.2} list some values of the coefficients $A^q_{lm}$ and $B^q_{lm}$, respectively. The transition weights $A^q_{lm}$ increase with increasing $m$ for $q=+1$ and decrease for $q=-1$, while the coefficients $B^q_{lm}$ exhibit the opposite behavior. This reflects the number of allowed values of the new $z$-component of the orbital angular momentum, ($m'=m+q$), as it changes during a transition ($l\rightarrow l\pm 1$).

On the other hand, from Eq.~(\ref{sigmank}) and the selection rules given in Eq.~(\ref{selection}), the branching ratio defined in Eq.~(\ref{branching}) satisfies
\beq
\frac{1}{n^2}\sum_{l=0}^{n-1}(2l+1)\left(\xi_{nl,n'l-1}+\xi_{nl,n'l+1} \right)=1.
\eeq
%

%=============================================
\begin{table}
\caption{Polynomials $Q_{nl,kl'}(k)$  as given by Eq. (\ref{fnlkl}) for $1\le n\le 5$.
\label{T.4}}
\setlength{\tabcolsep}{3pt}
\begin{small}
\begin{tabular}{cl}\hline\hline
 $nl$-$kl'$ & $\hskip.4in Q_{nl,kl'}(k)$ \\
\hline 
 $1s$-$kp$ & $1$ \\ & \\
 $2s$-$kp$ & $16 \left(1 + k^2\right) \left(4 + k^2\right)$ \\
 $2p$-$kd$ & $128 k^2 \left(1 + k^2\right)/9$ \\
 $2p$-$ks$ & $4 k^2 \left(4 + k^2\right)/9$ \\ & \\
 $3s$-$kp$ & $9 \left(1 + k^2\right) \left(9 + k^2\right) \left(27 + 7 k^2\right)^2$ \\
 $3p$-$kd$ & $432 k^2 \left(1 + k^2\right) \left(4 + k^2\right) \left(9 + k^2\right)$ \\
 $3d$-$kf$ & $7776 k^4 \left(1 + k^2\right) \left(4 + k^2\right)/25$  \\
 $3p$-$ks$ & $24 k^2 \left(3 + k^2\right)^2 \left(9 + k^2\right)$ \\
 $3d$-$kp$ & $144 k^4 \left(1 + k^2\right) \left(9 + k^2\right)/25$ \\ & \\
 $4s$-$kp$ & $256 \left(1 + k^2\right) \left(16 + k^2\right) \left(768 + 288 k^2 + 23 k^4\right)^2/9$ \\
 $4p$-$kd$ & $8192 k^2 \left(1 + k^2\right) \left(4 + k^2\right) \left(16 + k^2\right) \left(80 + 9 k^2\right)^2/45$ \\
 $4d$-$kf$ & $1048576 k^4 \left(1 + k^2\right) \left(4 + k^2\right) \left(9 + k^2\right) \left(16 + k^2\right)/75$ \\
 $4f$-$kg$ & $16777216 k^6 \left(1 + k^2\right) \left(4 + k^2\right) \left(9 + k^2\right)/2205$ \\  
 $4p$-$ks$ & $16 k^2 \left(16 + k^2\right) \left(1280 + 608 k^2 + 57 k^4\right)^2/45$ \\
 $4d$-$kp$ & $2048 k^4 \left(1 + k^2\right) \left(16 + k^2\right) \left(48 + 7 k^2\right)^2/225$ \\
 $4f$-$kd$ & $65536 k^6 \left(1 + k^2\right) \left(4 + k^2\right) \left(16 + k^2\right)/735$ \\ & \\
 $5s$-$kp$ & $625 \left(1 + k^2\right) \left(25 + k^2\right) \left(46875 + 20625 k^2 + 2545 k^4 + 91 k^6\right)^2/9$ \\
 $5p$-$kd$ & $10^4 k^2 \left(1 + k^2\right) \left(4 + k^2\right) \left(25 + k^2\right) \left(9375 + 1650 k^2 + 67 k^4\right)^2/81$ \\
 $5d$-$kf$ & $10^5 k^4 \left(1 + k^2\right) \left(4 + k^2\right) \left(9 + k^2\right) \left(25 + k^2\right) \left(175 + 11 k^2\right)^2/21$ \\
 $5f$-$kg$ & $2\times10^8 k^6 \left(1 + k^2\right) \left(4 + k^2\right) \left(9 + k^2\right) \left(16 + k^2\right) \left(25 + k^2\right)/441$ \\
 $5g$-$kh$ & $10^9 k^8 \left(1 + k^2\right) \left(4 + k^2\right) \left(9 + k^2\right) \left(16 + k^2\right)/5103$ \\
 $5p$-$ks$ & $200 k^2 \left(25 + k^2\right) \left(46875 + 25875 k^2 + 3725 k^4 + 149 k^6\right)^2/81$ \\
 $5d$-$kp$ & $2000 k^4 \left(1 + k^2\right) \left(25 + k^2\right) \left(2625 + 590 k^2 + 29 k^4\right)^2/63$ \\
 $5f$-$kd$ & $320000 k^6 \left(1 + k^2\right) \left(4 + k^2\right) \left(25 + k^2\right) \left(25 + 2 k^2\right)^2/147$ \\
 $5g$-$kf$ & $8\times 10^6 k^8 \left(1 + k^2\right) \left(4 + k^2\right) \left(9 + k^2\right) \left(25 + k^2\right)/5103$ \\
% &\txr{---}\\
\hline\hline
\end{tabular}
\end{small}
%\tablefoot{...}
\end{table}
%=============================================

%=============================================
\begin{table*}
\caption{Polynomials $Q_{nl,kl'}(k)$ as given by Eq. (\ref{fnlkl}) for $n=6,7$.
\label{T.4b}}
\setlength{\tabcolsep}{3pt}
\begin{small}
\begin{tabular}{cl}\hline\hline
 $nl$-$kl'$ & $\hskip2.1in Q_{nl,kl'}(k)$ \\
\hline 
 $6s$-$kp$ & $36\left(1 + k^2\right) \left(36 + k^2\right) \left(50388480
+ 24261120 k^2 +3654720 k^4 +211104 k^6 +4046 k^8 \right)^2/25$ \\
 $6p$-$kd$ & $384 k^2\left(1 + k^2\right)\left(4 + k^2\right)\left(36 + k^2\right) \left(4898880 +1061424k^2 +70092k^4+1425 k^6 \right)^2/35 $ \\
 $6d$-$kf$ & $ 2^{15}3^5 k^4\left(1 + k^2\right)\left(4 +k^2\right)\left(9 +k^2\right)\left(36 + k^2\right) \left(9072 +936k^2 +23 k^4 \right)^2/175 $ \\
 $6f$-$kg$ & $  2^{17}3^5 k^6\left(1 + k^2\right)\left(4 +k^2\right)\left(9 +k^2\right)\left(16 + k^2\right)\left(36 + k^2\right) \left(324+ 13 k^2 \right)^2/245 $ \\
 $6g$-$kh$ & $ 2^{21}3^5 k^8\left(1 + k^2\right)\left(4 +k^2\right)\left(9 +k^2\right)\left(16 +k^2\right)\left(25 +k^2\right)\left(36 +k^2\right)/35$\\
 $6h$-$ki$ & $  2^{24}3^8 k^{10}\left(1 + k^2\right)\left(4 +k^2\right)\left(9 +k^2\right)\left(16 + k^2\right)\left(25 + k^2\right) /21175 $ \\
 $6p$-$ks$ & $ 12 k^2\left(36 + k^2\right) \left( 19595520 +11757312 k^2 +2057184 k^4 +132528 k^6 +2761 k^8 \right)^2/35 $ \\
 $6d$-$kp$ & $ 2^{10}3^2 k^4 \left(1 + k^2\right)\left(36 + k^2\right) \left( 326592 +89424 k^2 + 7092 k^4 +167 k^6 \right)^2/175 $ \\
 $6f$-$kd$ & $ 2^{13}3^6 k^6 \left(1 + k^2\right)\left(4 + k^2\right)\left(36 + k^2\right) \left( 1296 +168 k^2 + 5 k^4 \right)^2/245 $ \\
 $6g$-$kf$ & $ 2^{17}3^5 k^8 \left(1 + k^2\right)\left(4 + k^2\right)\left(9 + k^2\right) \left(36 + k^2\right) \left(20+ k^2 \right)^2/175 $ \\
 $6h$-$kg$ & $ 2^{19}3^5  k^{10} \left(1 + k^2\right)\left(4 + k^2\right)\left(9 + k^2\right)\left(16 + k^2\right)\left(36 + k^2\right) /4235 $ \\
& \\
 $7s$-$kp$ & $7^4 \left(1 +k^2\right) \left(49 +k^2\right) \left(12711386205 +6485401125 k^2 +1097665170 k^4 +79704282 k^6 +2547265k^88 +29233 k^{10} \right)^2/2025$ \\
 $7p$-$kd$ & $2^5 7^6 k^2\left(1 +k^2\right)\left(4 +k^2\right) \left(49 + k^2\right) \left(86472015 +21176820 k^2 +1766450 k^4 +60116 k^6 +711 k^8 \right)^2/2025$ \\
 $7d$-$kf$ & $2^6 7^6 k^4\left(1 +k^2\right)\left(4 +k^2\right)\left(9 +k^2\right) \left(49 +k^2\right) \left(7411887+ 972405 k^2 +40229 k^4 +527 k^6 \right)^2/2025$ \\
 $7f$-$kg$ & $2^9 7^7 k^6\left(1 +k^2\right)\left(4 +k^2\right)\left(9 +k^2\right)\left(16 +k^2\right) \left(49 +k^2\right) \left(108045 +7350 k^2 +121 k^4\right)^2/6075$ \\
 $7g$-$kh$ & $2^9 7^{10} k^8\left(1 +k^2\right)\left(4 +k^2\right)\left(9 +k^2\right)\left(16 +k^2\right)\left(25 +k^2\right) \left(49 +k^2\right) \left(539 +15 k^2 \right)^2/40095$ \\
 $7h$-$ki$ & $2^{13} 7^{12} k^{10}\left(1 +k^2\right)\left(4 +k^2\right)\left(9 +k^2\right)\left(16 +k^2\right)\left(25 +k^2\right)\left(36 +k^2\right) \left(49 +k^2\right)/245025$ \\
 $7i$-$kk$ & $2^{14} 7^{13} k^{12}\left(1 +k^2\right)\left(4 +k^2\right)\left(9 +k^2\right)\left(16 +k^2\right)\left(25 +k^2\right)\left(36 +k^2\right)/11293425$ \\
 $7p$-$ks$ & $2^4 7^2 k^2\left(49 + k^2\right) \left(4237128735 +2680632465 k^2 +525218750 k^4 +42435274 k^6 +1471715 k^8 +18021 k^{10} \right)^2/2025 $ \\
 $7d$-$kp$ & $2^5 7^4 k^4 \left(1 +k^2\right) \left(49 + k^2\right)\left(155649627 +47799108 k^2 +4760154 k^4 +186788 k^6 +2483 k^8\right)^2/6075$ \\
 $7f$-$kd$ & $2^9 7^5 k^6 \left(1 +k^2\right)\left(4 +k^2\right)\left(49 +k^2\right) \left(21+ k^2 \right)^2 \left(36015 +4165 k^2 +94 k^4 \right)^2/2025$ \\
 $7g$-$kf$ & $2^9 7^6 k^8 \left(1 +k^2\right)\left(4 +k^2\right)\left(9 +k^2\right) \left(49 +k^2\right) \left(132055+ 11074 k^2 + 219 k^4\right)^2/200475$ \\
 $7h$-$kg$ & $2^{12} 7^8 k^{10} \left(1 +k^2\right)\left(4 +k^2\right)\left(9 +k^2\right)\left(16 +k^2\right)\left(49 + k^2\right)\left(147 +5 k^2\right)^2/147015$ \\
 $7i$-$kh$ & $2^{13}7^{10} k^{12}\left(1 +k^2\right)\left(4 +k^2\right)\left(9 +k^2\right)\left(16 +k^2\right)\left(25 +k^2\right)\left(49 +k^2\right)/11293425$ \\
\hline\hline
\end{tabular}
\end{small}
%\tablefoot{...}
\end{table*}
%=============================================

%=============================================
\begin{table*}
\caption{Polynomials $P_n(k)$ in Eq. (\ref{PQ}).
\label{T.3}}
\setlength{\tabcolsep}{3pt}
\begin{small}
\begin{tabular}{cl}
\hline
 $n$ & $\hskip2.5in P_{n}(k)$ \\
\hline
 1  & 1  \\
 2  & $\left(4 +3 k^2 \right) \left(4 +5 k^2\right)$ \\
 3  & $\left(81 +78 k^2 +13 k^4 \right) \left(81 +126 k^2 +29 k^4\right)$  \\
 4  & $\left(12288 +13056 k^2 +3152 k^4 +197 k^6 \right)
      \left(12288 +20736 k^2 +6800 k^4 +539 k^6\right)/9  $ \\
 5  & $\left(1171875 +1312500 k^2 +372250 k^4 +36100 k^6 +1083 k^8\right) 
      \left(1171875 +2062500 k^2 +786250 k^4 +95700 k^6 +3467 k^8 \right)/9 $ \\
 6  & $\left(302330880 +349920000 k^2 +108708480 k^4 
      +12909024 k^6 +628260 k^8 +10471 k^{10}\right)\times $ \\
    & $\left( 302330880 +545875200 k^2 +226281600 k^4 +33480864 k^6
       +1953540 k^8 +38081 k^{10}\right)/25 $ \\
 7  & $\left(622857924045 +737260399890 k^2 +242899890135 k^4 +32480603148 k^6 
       +1993432651 k^8 +55606082 k^{10} +567409 k^{12}\right)\times  $ \\
    & $\left(622857924045 +1144024758450 k^2 +500240606775 k^4 +82901603148 k^6
       +6067218955 k^8 +196890722 k^{10} +2297425 k^{12}\right)/2025 $\\
\hline\hline
\end{tabular}
\end{small}
%\tablefoot{...}
\end{table*}
%=============================================

%%%%%%%%%%%%%%%%%%%%%%%%%%%%%%%%%%%%%%%%%%%%%%%%%%%%%%%%%%%%%%%%%
\section{Branching fractions}\label{s:branching}

For the evaluation of the branching fractions [Eq.~(\ref{branching})], the partial cross section $\sigma_{nl,kl'}$ can be conveniently expressed in terms of the oscillator strength \citep{bethe1957, burgess1965},
\beq\label{sigma_fosc}
\sigma_{nl,kl'} = 4\pi^2 \alpha a_0^2 \frac{df_{nl,kl'}}{d\cm{E}},
\eeq
with $\alpha=e^2/(\hbar c)$ the fine-structure constant, $a_0=\hbar^2/(e^2 m_\text{e})$ the Bohr radius, and $df_{nl,kl'}/d\cm{E}$ the oscillator-strength density in the continuum. Here, $\cm{E}$ is measured in Rydbergs.
A convenient method for obtaining the oscillator-strength density is provided by the analytic continuation approach introduced by \citet{menzel1935}. This method shows that the oscillator strength  $f_{nl,n'l'}$ for transitions between discrete states (from $nl$ to $n'l'$) can be extended into the continuum according to
\beq\label{continuity}
\frac{df_{nl,kl'}}{d\cm{E}} =\left[\frac{dn'}{d\cm{E}} f_{nl,n'l'} \right]_{n'=ik}
 = \left[\frac{n'^3}{2}f_{nl,n'l'}\right]_{n'=ik},
\eeq
where the energies of the sublevels $nl$, $n'l'$ and $kl'$ are given by
\beq\label{energy0}
\cm{E}_{n}=-n^{-2},\hskip.2in
\cm{E}_{n'}=-n'^{-2},\hskip.2in
\cm{E}_{k}= k^{-2}.
\eeq
Equation~(\ref{continuity}) involves an extrapolation of the transition energy $\cm{E}=\cm{E}_{n'}-\cm{E}_{n}$ into the continuum domain ($\cm{E}>-\cm{E}_n$, i.e., above the ionization threshold from level $n$), with the identification  
\beq
n'=ik,\hskip.3in n'^2=-k^2.
\eeq
A formal proof of the Menzel-Pekeris method was provided by \citet{grant1958} in the context of free-free transitions.
%
%=============================================
\begin{figure*}
\includegraphics[width=1.\textwidth]{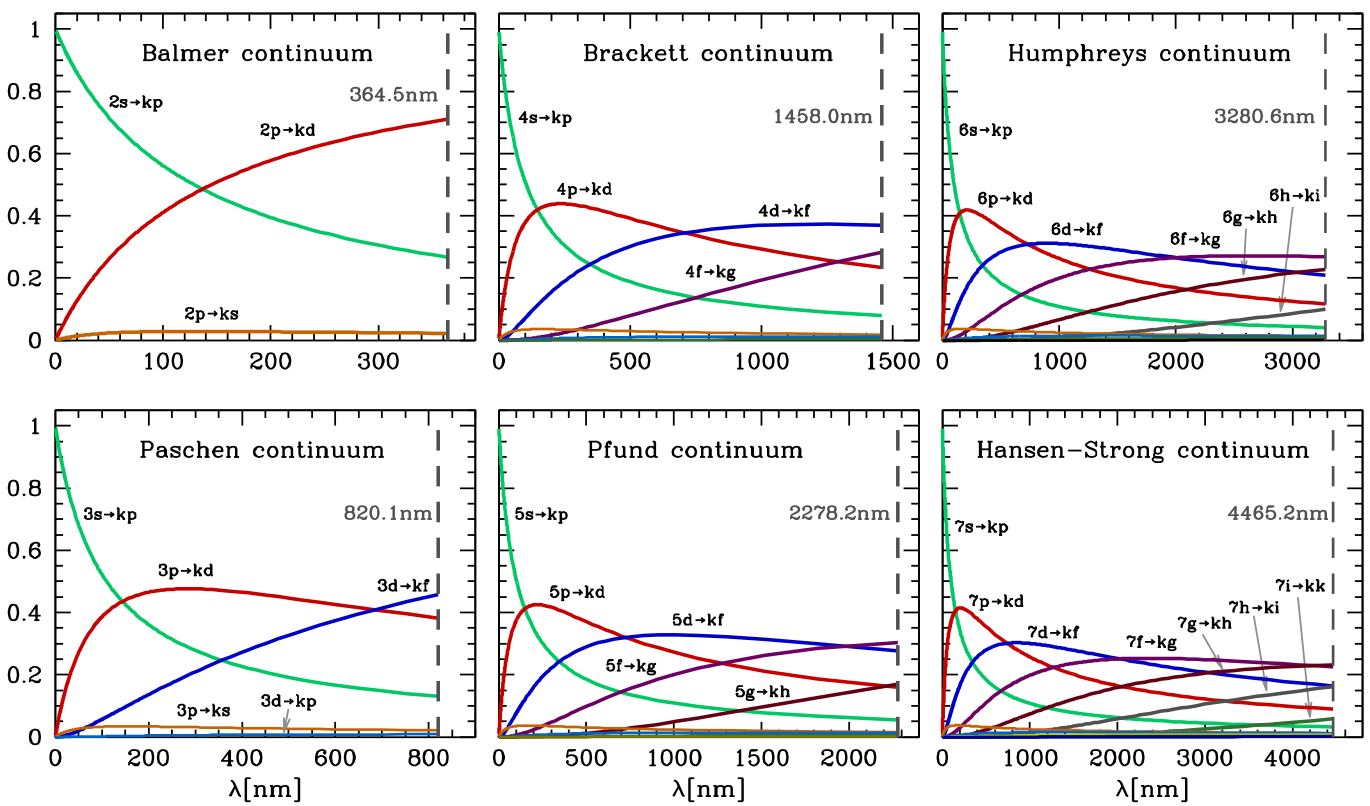}
\caption{Branching fractions in the form $(2l+1)n^{-2}Q_{nl,kl'}/P_{n,k}$ 
as a function of the light wavelength, for continua from Balmer ($n=2$) to Hansen-Strong ($n=7$) in zero magnetic field. Vertical dashed line denotes the limit wavelength of photoionization. The strongest transitions are labeled.}
\label{f:branch}
\end{figure*}
%=============================================

The oscillator strength for transitions from a lower state $nl$ to an upper state $n'l'$ is given by \citep{bethe1957}
\beq\label{fnlnl}
f_{nl,n'l'}= \frac{l_>}{3(2l+1)} \cm{E}_{nl,n'l'}
  \left[\int_0^\infty \frac{r^3}{a_0} R_{n'l'}(r) 
 R_{nl}(r)dr\right]^2,
\eeq
where $l_>$ denotes the larger of $l$ and $l'$, $\cm{E}_{nl,n'l'}$ is the transition energy, and $R_{nl}(r)$ is the normalized radial wavefunction of the $nl$ state,
\beq\label{radial}
R_{nl}(r)= \frac{2}{n^2a_0^{3/2}}\left(\frac{(n-l-1)!}{[(n+l)!]^3} \right)^{1/2}
\left( \frac{2Zr}{na_0} \right)^l \exp\left( -\frac{Zr}{na_0}\right) 
L_{n-l-1}^{2l+1} \left(\frac{2Zr}{na_0} \right),
\eeq 
with $L_a^b(z)$ the associated Laguerre polynomial,
\beq
L_a^b(z)= \sum_{k=0}^a (-1)^{k} \frac{[(a+b)!]^2}{(a-k)!(b+k)!k!} z^k .
\eeq 
Solutions of Eq. (\ref{fnlnl}) take the form
\beq\label{fnlnlex}
 f_{nl,n'l'}=\frac{256}{3}\; n^3n'^5\; \tilde{Q}_{nl,n'l'}\; 
             \frac{(n'-n)^{2(n'-n-1)}}{(n'+n)^{2(n'+n+1)}}.
\eeq
Here, $\tilde{Q}_{nl,n'l'}$ is a polynomial in $n'^2$ of order $2(n-1)$. According to the selection rules given in Eq.~(\ref{selection}), only terms with $l'=l\pm 1$ yield nonzero values of $\tilde{Q}_{nl,n'l'}$.
Applying Eq.~(\ref{continuity}) then yields
\beq\label{fnlkl}
\frac{df_{nl,kl'}}{d\epsilon}=\frac{128}{3}\; \frac{n^3k^8 Q_{nl,kl'}}
 {(n^2+k^2)^{2n+2}}   \;
 \frac{\exp\left[-4k\arctan(n/k)\right]}{1-\exp(-2\pi k)},
\eeq
with 
\beq
Q_{nl,kl'} = \tilde{Q}_{nl,n'l'}|_{n'=ik}.
\eeq
This procedure was used to derive all coefficients $Q_{nl,kl'}$ for $1\le n\le 7$. Their explicit expressions, listed in Tables~\ref{T.4} and \ref{T.4b}, correct several erroneous forms previously reported in the literature, most notably for the $n=7$ case in \citet{hatanaka1946}. Because the coefficients of $\tilde{Q}_{nl,n'l'}$ are negative for odd powers of $n'^2$ and positive for even powers, the resulting functions $Q_{nl,kl'}$ are polynomials in $k^2$ with strictly positive coefficients.

The total cross section $\sigma_{n,k}$ is then obtained using Eqs.~(\ref{sigmank}) and (\ref{sigma_fosc}),
\beq\label{sigmatotal}
\sigma_{n,k} = 4\pi^2 \alpha a_0^2 \frac{df_{n,k}}{d\cm{E}},
\eeq
which yields
\beq\label{fnk}
\frac{df_{n,k}}{d\epsilon}=\frac{128}{3} \; \frac{n^3k^8 P_{nk}}
 {(n^2+k^2)^{2n+2}} \;
  \frac{\exp\left[-4k\arctan(n/k)\right]}{1-\exp(-2\pi k)} ,
\eeq
with
\beq\label{PQ}
P_{nk} =\frac{1}{n^2}\sum_{l=0}^{n-1} (2l+1)
  \left( Q_{nl,kl-1} + Q_{nl,kl+1}  \right).
\eeq
The first seven polynomials  $P_{nk}$ ($1\le n\le 7$) are listed in Table~\ref{T.3}. Each polynomial $P_{nk}$ is related through
\beq\label{tildePnn}
P_{n,k} = \tilde{P}_{n,n'}\hskip.3in (n'=ik)
\eeq
where the polynomial $\tilde{P}_{n,n'}$ is associated with the mean bound–bound oscillator strength $f_{n,n'}$,
\beq\label{fnlnlex}
 f_{n,n'}=\frac{256}{3}\; n^3n'^5\; \tilde{P}_{n,n'}\; 
             \frac{(n'-n)^{2(n'-n-1)}}{(n'+n)^{2(n'+n+1)}}.
\eeq
The present results, through Eq.~(\ref{tildePnn}), correct two minor errors in $\tilde{P}_{7,n'}$ reported by \citet{menzel1935} (see their Eq.~1.28), and also supply the expression for $\tilde{P}_{6,n'}$, which was omitted in that work.
% (5606082 instead 55606082 in 10 degree coefficient and 32628840888 instead 32480603148 in 6 degree coefficient)

Finally, the evaluation of the branching fractions reduces to the simple ratio
\beq\label{xi}
\xi_{nl,kl'}\equiv\frac{Q_{nl,kl'}}{P_{nk}}.
\eeq
Figure~\ref{f:branch} illustrates the wavelength dependence of the branching fractions for photoionization from substates with $2\le n\le 7$.

The complexity of the polynomials $P_{n,k}$ and $Q_{nl,kl'}$ increases rapidly at high principal quantum numbers, making a more efficient evaluation scheme desirable. As shown in Fig.~\ref{f:branch}, transitions with $l\rightarrow l+1$ dominate the photoionization strength, while those with $l\rightarrow l-1$ progressively weaken,
\beq \label{xis}
\xi_{nl,k\,l+1} \gg \xi_{nl,k\,l-1},\hskip.4in n\gg 1.
\eeq
On the other hand, the cross section for the sublevels $nl$ can be evaluated using the approximation
\beq \label{kramers}
\sigma_{nl,k}=\left(\xi_{nl,k\,l+1} + \xi_{nl,k\,l-1}\right)\sigma_{n,k}
 = \sigma_{n,k}^\text{Kramers} g_{nl},
\eeq
where $\sigma_{n,k}^\text{Kramers}$ is the Kramers cross section for transitions $n\rightarrow k$, and $g_{nl}$ is the bound–free Gaunt factor associated with the sublevels $(n,l)$. Similarly, the mean cross section $\sigma_{n,k}$ can be expressed in terms of $l$-averaged values of the Gaunt factor $g_{nl}$
\beq \label{kramers2}
\sigma_{n,k}= \sigma_{n,k}^\text{Kramers} g_{n},\hskip.3in
g_n= \frac{1}{n^2}\sum_{l=0}^{n-1}(2l+1)g_{nl}.
\eeq
Taking into account Eqs.~(\ref{xis})-(\ref{kramers2}), for $n\ge 8$ we adopt the approximations 
\beq
\xi_{nl,k\,l+1} \approx \frac{ g_{nl}}{g_{n}},
\hskip.5in \xi_{nl,k\,l-1} \approx 0.
\eeq
The bound–free Gaunt factors are calculated using the analytical expressions provided by \citet{rozsnyai1988}
\beq
g_{nl}=\begin{cases}a_1\omega(a_2 +\omega)^{-a_3}& 1\le \omega\le \omega_m, \\
a_4\omega(a_5 +\omega)^{-l-3/2} & \omega_m < \omega <\infty ,
\end{cases}
\eeq
with constants $a_i$ and $\omega_m$ specified in that work.
The Kramers cross section $\sigma_{n,k}^\text{Kramers}$ is given by Eq.  (\ref{sigmatotal}) with
\beq \label{kr.fkn}
\frac{df_{n,k}}{d\cm{E}} \approx
\frac{16}{3\sqrt{3}\pi} \frac{1}{n^5}\left(\frac1{n^2}+\frac1{k^2}\right)^{-3}.
\eeq
%

%%%%%%%%%%%%%%%%%%%%%%%%%%%%%%%%%%%%%%%%%%%%%%%%%%%%%%%%%%%%%%%%%
\section{Opacity in presence of magnetic fields}\label{s:rigid}

The magnetic field in an MWD atmosphere breaks the degeneracy of each energy level $n$, splitting it into multiple sublevels $\xi$. As a consequence, the absorption coefficient $\chi^q$ (in units of cm$^{-1}$) for photons of energy $\cm{E}$ and polarization $q$ is composed of contributions from many different initial atomic states,
\beq
\chi^q (\cm{E})= \sum_\xi \sigma_{\xi}^q(\cm{E}) n_\xi,
\eeq
where $n_\xi$[cm$^{-3}$] is the occupation number of the sublevel $\xi$. 

By assuming that the bound and free electron wavefunctions in the vicinity of the atom remain unperturbed by the magnetic field, so that the transition matrix element in Eq.~(\ref{basic}) is unchanged, the cross section within the RWA framework can be written as
\beq\label{eRWA}
\sigma_{\xi}^q(\cm{E})=\frac{\cm{E}}{\cm{E} -\Delta_\xi} 
                          \sigma_\xi^{q,0}(\cm{E} -\Delta_\xi),
\eeq
where $\sigma_\xi^{q,0}$ is the zero-field cross section and $\Delta_\xi$ is the shift of the ionization threshold induced by the field,
\beq\label{delta}
\Delta_\xi \equiv \cm{E}_{\xi,*}-\cm{E}_{\xi,0},
\eeq
with $\cm{E}_{\xi,*}$ and $\cm{E}_{\xi,0}$ denoting the ionization thresholds in the presence and absence of the magnetic field, respectively. In practice, Eq.~(\ref{eRWA}) corresponds to a rescaling of the cross-section amplitude and to a modification of the quasi-principal quantum number $k$ at which Eq.~(\ref{e.main}) is evaluated. As indicated by Eq.~(\ref{energy0}), $k$ depends on the excess of photon energy above the ionization threshold, such that
\beq\label{kE}
k= (\cm{E}-\cm{E}_{\xi,*})^{-1/2}.
\eeq
%

%%%%%%%%%%%%%%%%%%%%%%%%%%%%%%%%%%%%%%%%%%%%%%%%%%%%%%%%%%%%%%%%%
\subsection{Eigenenergies}\label{s:energy}

There are several ways to identify the Hamiltonian eigenstates of magnetized atoms. In the limit of low field strength (Zeeman approximation), $\beta\ll1$ ($\beta= B/B_0$, $B_0=2m_\text{e}^2ce^3/\hbar^3 \approx 4.70103\times 10^9$~G), the atomic states are labeled by the usual quantum numbers $\xi=\{n,l,m,m_s\}$ corresponding to the field-free atom ($m_s=\pm\frac12$ denotes the $z$-component of the electron spin).
For hydrogen atoms at rest in a magnetic field, the binding energies $\cm{E}_{\xi}$, including linear and quadratic Zeeman corrections (i.e., exact to second order in $\beta$; see \citealt{guth1929, vanvleck1932, schiff1939, garstang1977}), are given by
\bea\label{zeeman}
\cm{E}_{\xi}&=&-\frac{1}{n^2} +4\left(\frac{m}{2}+m_s\right)\beta \cr
 &&  +\frac{n^2[5n^2+1-3l(l+1)](l^2+l-1+m^2)}{(2l-1)(2l+3)} \beta^2.
\eea
At the opposite extreme, for very strong fields ($\beta\gg1$), the eigenstates have been studied using the adiabatic approximation \citep{schiff1939, canuto1972} and are classified by the set $\xi=\{N,\nu,m,m_s\}$. Here, $N$ ($=0,1,2,\dots$) denotes the quantization of the electron motion perpendicular to the field (Landau levels), while $\nu$ specifies the wavefunction excitations along the field.

The states in intense magnetic fields and those of field-free atoms are connected through the noncrossing rule \citep{simola1978}, where the quantum numbers $m$, $m_s$, and the $z$-parity $\pi_z=\pm1$ (reflection symmetry along the field direction) are conserved owing to the symmetries of the potential energy. The remaining quantum numbers,  $(n,l)$ and $(N,\nu)$, are related for bound states by expressions that preserve the ordering of increasing energy in the limits $\beta\rightarrow 0$  and $\beta\rightarrow\infty$ \citep{vera2020}. As a result, there is a one-to-one correspondence between the two sets  $\{n,l,m,m_s\}$ and $\{N,\nu,m,m_s\}$, and either labeling scheme may be used.

Exact binding energies of hydrogen atoms at rest for arbitrary magnetic field strengths can be computed numerically \citep{kravchenko1996}. In this work, we employ analytical fits \citep{vera2020} to accurate numerical calculations \citep{schimeczek2014}. These fits cover an arbitrary number of atomic states and are supplemented by Eq.~(\ref{zeeman}) to ensure high accuracy in the low-field limit $\beta\rightarrow 0$. The energy evaluations include center-of-mass corrections arising from internal motion \citep{pavlov1980} and collective motion \citep{vincke1988}. The latter accounts for the coupling between the binding energy of a magnetized atom and its motion perpendicular to the field.

For the magnetic-field regime considered in this work  ($\beta\ll 1$), the total energy of an atom may be written as
\beq \label{e.EE}
E = \cm{E}_\xi +\frac{k_\perp^2}{2M_{\perp,\xi}} +\frac{k_z^2}{2M_\text{H}},
\eeq
where $\cm{E}_\xi$ is the energy of the atom at rest, $k_\perp$ and $k_z$ are the components of the pseudomomentum used to separate the center-of-mass motion from the relative proton–electron motion \citep{gorkov1968}, $M_\text{H}$ is the total mass of the hydrogen atom, and $M_{\perp,\xi}$ is its effective mass for motion perpendicular to the field \citep{pavlov1993}.

In the present work, we omit the so-called decentered atomic states. These are bound states in which the electron wavefunction is displaced from the Coulomb potential well to a magnetic well \citep{burkova1976} and, as such, cannot be treated within the RWA framework. The formation of these states requires both high magnetic field strengths and large transverse pseudomomenta.
Preliminary estimates, based on expressions originally derived for neutron star atmospheres and adapted to MWD conditions \citep{vera2020}, indicate that decentered states may occur at the surfaces of the strong MWDs ($B\approx 470$~MG). Both approximations employed here --the RWA and the effective transverse mass used in Eq.~(\ref{e.EE}) corresponding to centered states-- are therefore applicable in weaker magnetic fields, where decentered states can be safely neglected.

The binding energy of an arbitrary bound state $\xi$ can be expressed as
\beq\label{bound}
\cm{E}_\xi=\cm{E}_{-|m|} + 2\beta\left(|m|+m+2m_s+1\right),
\eeq
where $\cm{E}_{-|m|}$ denotes the binding energy of the spin-down state with magnetic quantum number equal to $-|m|$.

For bound–free transitions $\xi\rightarrow \xi'$, the threshold energy of continuum states is
\beq\label{free}
\cm{E}_c= 2\beta\left(|m'|+m'+2m'_s+1\right),
\eeq
which corresponds to a free electron with no radial excitation and zero kinetic energy along the field \citep{rohrmann2025}.
The difference $\cm{E}_c-\cm{E}_\xi$ defines the ionization threshold required in Eqs.~(\ref{delta}) and (\ref{kE}). Taking into account that, for an electric-dipole transition, the magnetic quantum number changes as $m'=m+q$ and that the electron spin remains unchanged ($m_s'=m_s$), one obtains
\beq\label{ediff}
\cm{E}_c-\cm{E}_\xi = \begin{cases} 
-\cm{E}_{-|m|} + 4\beta, & q=+1,~m\ge 0, \\ 
-\cm{E}_{-|m|} - 4\beta, & q=-1,~m \ge 1, \\ 
-\cm{E}_{-|m|}, &  \text{otherwise}. \\
\end{cases}
\eeq
Because some initial states $\xi$ may lie above the continuum edge for a given polarization $q$ (particularly for $q=-1$ and $m\ge 1$), the minimum energy required for photoionization is
\beq\label{ebinding}
\cm{E}_{\xi,*}=\text{max}(0,\cm{E}_c-\cm{E}_\xi).
\eeq
In all cases, $\cm{E}_{\xi,*}\rightarrow \cm{E}_{\xi,0}=n^{-2}$ in the limit $\beta\rightarrow 0$, as expected.

%%%%%%%%%%%%%%%%%%%%%%%%%%%%%%%%%%%%%%%%%%%%%%%%%%%%%%%%%%%%%%%%%
\subsection{Occupation numbers}\label{s:occu}

The abundances of atoms ($n_\text{H}$), protons ($n_\text{p}$), and free electrons ($n_\text{e}$) were determined from the ionization balance expressed by the chemical-potential relation $\mu_\text{H}=\mu_\text{p}+\mu_\text{e}$, which yields (see \citealt{vera2020} for details)
\beq \label{e.Q}
\frac{n_\text{H}}{n_\text{e}n_\text{p}}
=\frac{\lambda_e^3}{2} f(\eta) Z_\text{H}. 
\eeq
Here, $\lambda_\text{e}$ is the electron thermal wavelength, $Z_\text{H}$ the atomic partition function, and $f(\eta)$ a factor accounting for the excess chemical potentials of free charges. This factor depends on $\eta=2\beta/(k_\text{B}T)$, where $k_\text{B}$ is the Boltzmann constant and $T$ the gas temperature (with thermal energy $k_\text{B}T$ expressed in Rydbergs). The function $f(\eta)$ arises from the interaction of charged particles with the magnetic field and from modifications of the density of states caused by the transverse motion being frozen into Landau orbitals.
Using Eq.~(\ref{e.EE}), the atomic partition function was evaluated as
\beq
Z_\text{H}=\sum_\xi \frac{M_{\perp,\xi}}{M_\text{H}}w_\xi e^{-\cm{E}_\xi/ k_\text{B}T},
\eeq
with the effective atomic mass ($M_{\perp,\xi}$) depending on the internal state and field strength, and $w_\xi$ being an occupation probability that accounts for density effects on the atom.
The occupation numbers of bound states were determined from
\beq\label{nxi}
n_\xi= \frac{n_\text{H}}{Z_\text{H}} 
\frac{M_{\perp,\xi}}{M_\text{H}} w_\xi e^{-\cm{E}_\xi/ k_\text{B}T}.
\eeq
Note that, under ionization equilibrium, Eq.~(\ref{nxi}) applies to all atomic states, irrespective of whether their energies lie below or above the photoionization continuum threshold, which itself depends on the photon polarization. 
Abundances of states with relatively high energies (e.g., metastable states) are expected to be very small.

%%%%%%%%%%%%%%%%%%%%%%%%%%%%%%%%%%%%%%%%%%%%%%%%%%%%%%%%%%%%%%%%%
\section{Results and discussion}\label{s:results}

%=============================================
\begin{figure}
\includegraphics[width=.5\textwidth]{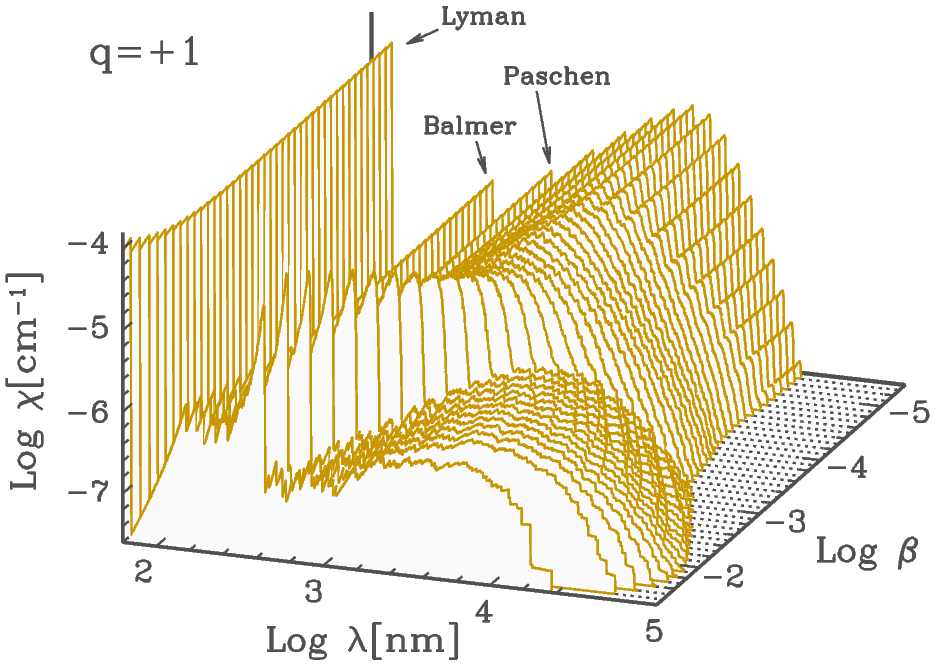}\vskip.2in
\includegraphics[width=.5\textwidth]{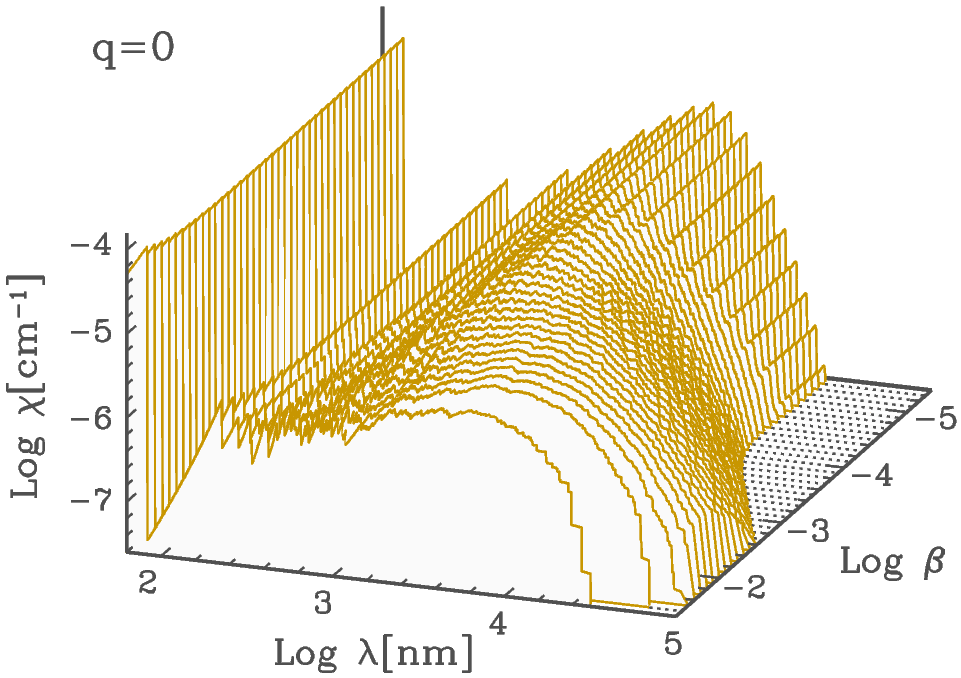}\vskip.2in
\includegraphics[width=.5\textwidth]{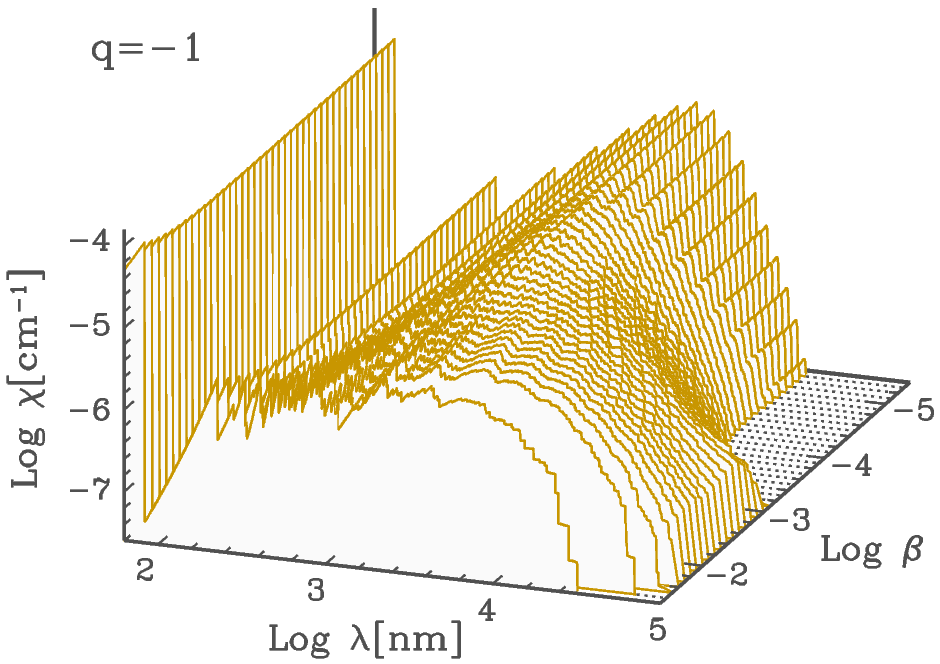}
\caption{Extinction coefficient due to photoionizations from atomic hydrogen at $T=20000$~K and $\log \rho=10^{-8}$~g/cm$^3$, calculated for various photon polarizations ($q=0,\pm1$) and different magnetic field strengths.}
\label{f:Xq}
\end{figure}
%=============================================

%=============================================
\begin{figure*}
\includegraphics[width=.333\textwidth]{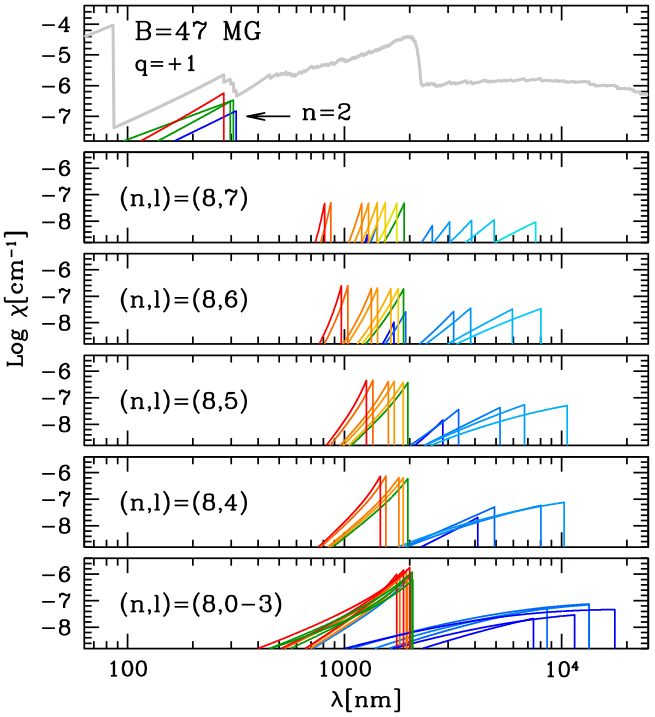}
\includegraphics[width=.32\textwidth]{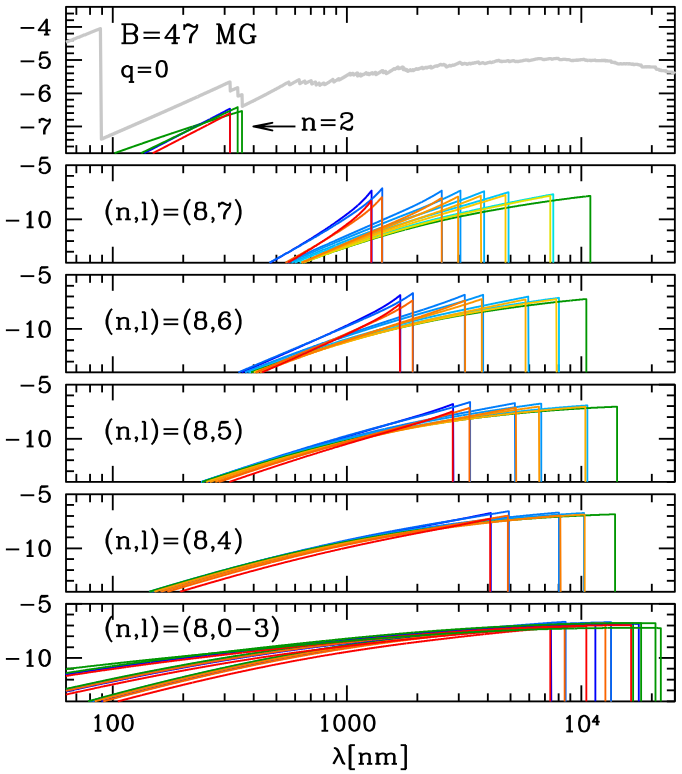}
\includegraphics[width=.32\textwidth]{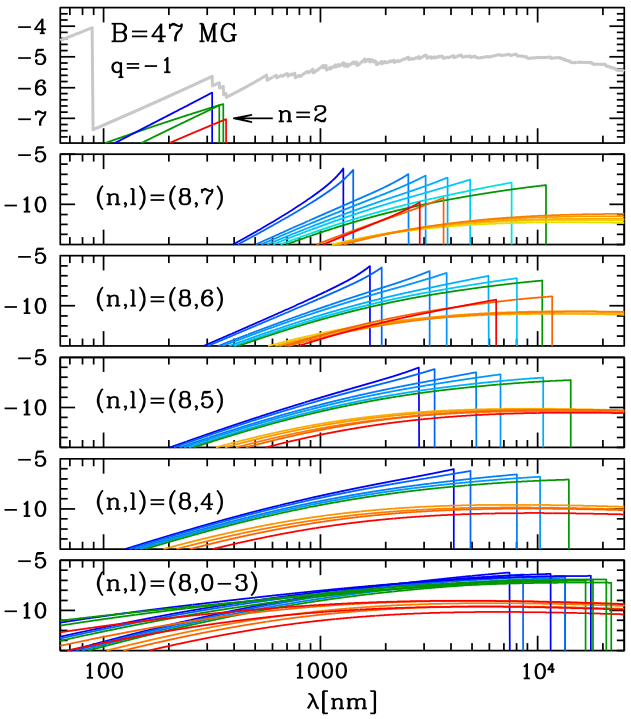}
\caption{Total photoionization absorption $\chi^q$ (gray thick line) and partial contributions ($\sigma^q_\xi n_\xi$) from spin-down sublevels at $n=2$ and $n=8$ manifolds calculated for $B\approx47$ MG ($\log\beta=-2$) and photon polarizations $q=0,\pm1$.
The colors distinguish the contributions of states with different values of $m$: green for $m=0$, light blue for $m<0$ (blue for $m=-l$), light red for $m>0$ (red for $m=l$).}
\label{f:Xqn8}
\end{figure*}
%=============================================
To illustrate the effects of the RWA on the photoionization opacity of magnetized hydrogen atoms, we consider a gas with temperature $T=20000$~K and density $\rho=10^{-8}$~g~cm$^{-3}$. Figure~\ref{f:Xq} shows the absorption coefficient for the three photon polarizations ($q=0,\pm1$) over the magnetic-field range $-5.4\le\log\beta \le -1.3$ (19~kG~$\le B\le 235$~MG).

At the lowest field strengths, the photoabsorption coefficient is identical for all polarizations and exhibits the characteristic sequence of jumps (from levels $n=1$ to 17 in Fig.~\ref{f:Xq}) of the field-free spectrum. As the magnetic field increases, the continua originating from higher-$n$ levels develop multiple smaller jumps at different wavelengths, which progressively smooth the absorption profile from low to high photon energies.
For field strengths above $B\approx 0.5$~MG ($\log\beta\ga-4$), the absorption extends toward longer wavelengths. This effect is most pronounced for left-handed circular polarization ($q=-1$), owing to contributions from states with positive $m$, whose ionization thresholds are lowered by $4\beta$ [Eq.~(\ref{ediff})].
When the field exceeds $B\approx 10$~MG ($\log\beta\ga-2.7$), right-handed circularly polarized photoionization ($q=+1$) develops a bulge at wavelengths shorter than the cyclotron resonance (photon energy $\epsilon=4\beta$), which at higher field strengths evolves into an abrupt jump just below the resonance wavelength \citep{rohrmann2025}. In contrast, the central bulk of the absorption coefficient decreases slightly and shifts toward longer wavelengths for polarizations $q=0$ and $q=-1$.

For all polarizations, the Lyman continuum remains essentially unchanged up to the highest magnetic fields considered here, except that its intensity begins to increase noticeably for $\log\beta\approx-2$ ($B\approx 47$~MG), and the $\chi^+$ component shifts to higher photon energies. At the strongest fields, the absorption coefficients decrease at long wavelengths. In this regime, the dominant contributions arise from states with positive $m$ and/or $m_s$, which approach the continuum and experience a marked reduction in their occupation numbers, thereby explaining the decline in absorption.

\begin{figure}
\includegraphics[width=.48\textwidth]{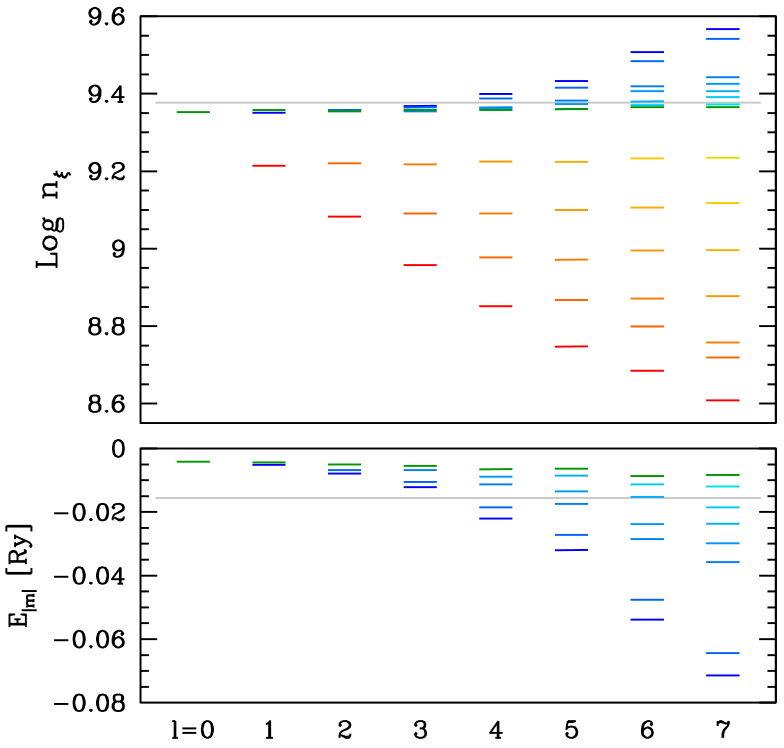}%\vskip.2in
\caption{Abundances and binding energies $\cm{E}_{-|m|}$ of spin-down atoms at sublevels associated to the $n=8$ manifold, ordered by the quantum number $l$. The colors distinguish the value of $m$: green for $m=0$, light blue for $m<0$ (blue for $m=-l$), light red for $m>0$ (red for $m=l$). Gray lines show the corresponding value at zero-field (the same value for all sublevels associated to the field-free level $n=8$).}
\label{f:n8E}
\end{figure}

The behavior described above can be better understood by analyzing the individual contributions from states associated with specific zero-field  $n$ manifolds. Figure~\ref{f:Xqn8} shows the components arising from the $n=2$ and $n=8$ manifolds for $\log\beta=-2$ ($B\approx 47$~MG) and for the three photon polarizations $q=0,\pm1$. As shown in the upper panels, the total photoionization coefficient (gray curve) still exhibits the Lyman continuum and a remnant of the Balmer continuum.
The Balmer spectrum, however, displays distinct ionization edges that depend on the initial atomic state (i.e., the magnetic quantum number) and on the polarization of the radiation, as first described by Jordan (1992) and Merani et al. (1995). Contributions from all other excited states are largely blended together. The states belonging to the $n=8$ manifold provide a representative example of this behavior. Their abundances and binding energies ($\cm{E}_{-|m|}$) are shown in Fig.~\ref{f:n8E} for spin-down states with selected values of $l$ and $m$.
%=============================================
%=============================================

Absorption edges are shifted in the energy spectrum according to Eq.~(\ref{ediff}) as a function of the magnetic field. For $q=0$ (linearly polarized radiation), states with $\pm m$ share the same ionization edge (middle panels of Fig.~\ref{f:Xqn8}), whose position is determined exclusively by the value of $\cm{E}_{-|m|}$. The ionization edges occur at higher energies for larger $|m|$ and display a broader distribution over the spectrum for states associated with high-$l$ manifolds, as inferred from the values shown in Fig.~\ref{f:n8E}.
By contrast, contributions from positive-$m$ states (light red curves) are additionally shifted by an amount $4\beta$ toward lower (higher) energies for radiation with  $q=-1$ ($+1$), as shown in Fig.~\ref{f:Xqn8}. As a consequence, photoionization by left-handed circularly polarized radiation extends mainly toward longer wavelengths, whereas right-handed circular polarization leads to an accumulation of contributions at wavelengths shorter than the cyclotron resonance.
This overlap near the cyclotron line becomes more pronounced as the field strength increases, increasingly involving states originating from higher-$n$ manifolds. This behavior gives rise to the abrupt jump mentioned above.\footnote{An approximate field strength at which positive-$m$ states from a given $n$ manifold contribute significantly to the $\chi^+$ feature can be estimated by comparing the absolute value of the zero-field binding energy ($1/n^2$) with the cyclotron energy $4\beta$, yielding $n>(4\beta)^{-1/2}$. As a guide, this corresponds to $n>9$, $5$, and $3$ for $B=14.5$, $47$, $130$~MG ($\log \beta=-2.5$, $-2.0$, $-1.6$), respectively.}

The total abundance of atoms increases with increasing $B$ (slowly at low field strengths) owing to the rise in the ionization energy. This energy reaches $-E_{1s,m_s=-1/2}=1.0191$~Ry at $B\approx 47$~MG. As a consequence, the occupation number of the ground state (1s$_{-1/2}$ in the zero-field notation) increases, whereas the populations of excited states mostly decrease. These changes are relatively moderate, but they become more pronounced above $B\sim 50$~MG.
Among all manifolds except $n=1$, the set of states associated with the zero-field $n=8$ manifold is the most populated under the adopted gas conditions, slightly exceeding the population of the $n=2$ manifold.
For fixed $n$ and  $l$, the binding energy -- and therefore the particle abundance -- decreases with increasing $m$ (Fig.~\ref{f:Xqn8}). In the ($n,l)=(8,7$) manifold, the population difference between the states with the smallest and largest values of $m$ reaches approximately one order of magnitude. This implies that, while the total population of a given $n$ manifold may not differ substantially from its zero-field value, the populations of its individual sublevels can vary dramatically, as shown in Fig.~\ref{f:n8E}.

The contribution of each transition $\sigma^q_\xi n_\xi$ to the total opacity depends on both the transition probability and the occupation number of the initial state. Their respective dependences on the magnetic quantum number help to explain the behavior shown in Fig.~\ref{f:Xqn8}. The cross sections $\sigma^q_\xi$ are dominated by $l\rightarrow l+1$ transitions, which involve the geometric weight factors $A^q_{lm}$ [Eq.~(\ref{e.main2})]. These factors decrease with increasing $m$ for left-handed circularly polarized light and increase for right-handed polarization.
For given values of $n$ and $l$, absorption of left-handed circularly polarized light decreases predominantly with increasing  $m$, because both the occupation number $n_\xi$ and the geometric factor $A^q_{lm}$ decrease (blue to red curves in the right panel of Fig.~\ref{f:Xqn8}). In addition, the corresponding absorption profiles are shifted toward longer wavelengths for transitions originating from states with $m>0$.
In contrast, for right-handed circularly polarized radiation, the decrease of $n_\xi$ with increasing $m$ is largely compensated by the increase of the geometric weight $A^q_{lm}$. This explains why photoionization from sparsely populated states ($m\ge 0$), whose transition energies are shifted to higher energies, produces the prominent $\chi^+$ feature near the cyclotron line (left panel of Fig.~\ref{f:Xqn8}).

%%%%%%%%%%%%%%%%%%%%%%%%%%%%%%%%%%%%%%%%%%%%%%%%%%%%%%%%%%%%%%%%%
\section{Concluding remarks}\label{s:concl}

We have presented a detailed procedure for evaluating the photoionization opacity in hydrogen atmospheres of MWDs with low and moderate magnetic field strengths. The method analyzed is an updated version of the rigid wavefunction approximation originally introduced by \citet{lamb1974}, complemented with accurate atomic energies for a theoretically unrestricted number of bound states at arbitrary magnetic field strengths \citep{vera2020}. The computation of the absorption coefficient also requires a self-consistent evaluation of the abundances of the initial bound states. To this end, we have calculated the ionization equilibrium of hydrogen in a magnetic field and obtained absolute photoionization opacities.

Because a magnetic field breaks the degeneracy of atomic energy levels, a large number of transitions contribute to the photoionization opacity in the field range $1\la B\la 100$~MG. Each photoionization process depends on the light polarization in two distinct ways: through a geometric probability factor (given by the Wigner $3j$ coefficients) and through the ionization threshold associated with a specific Landau level. Dichroic features arise mainly from the latter effect, since the continuum thresholds of positive-$m$ states are further shifted upward or downward by the cyclotron energy for right- or left-handed circularly polarized light, respectively. Photoionization from positive-$m$ states produces a strong jump in the continuum opacity (blueward of the first cyclotron resonance) for right-handed circular polarization and extends the opacity toward longer wavelengths for left-handed polarization. This anisotropy appears already at field strengths of a few MG and becomes increasingly pronounced at higher fields (Fig.~\ref{f:Xq}).

The total photoionization opacity is therefore strongly governed by the positions of the transition edges and by the abundances of the initial states, both of which are treated consistently and accurately within the present framework. A limitation of the RWA approach is that it does not reproduce the non-monotonic wavelength dependence of the cross sections obtained in fully rigorous calculations, although it reproduces their smoothed, mean behavior reasonably well. Given the very large number of transitions contributing to the total opacity, it is expected that fine details of accurate cross sections -- particularly resonance structures -- will be blurred by the summation over components and by the field spread when applied to MWD model atmospheres. This expectation can be tested once precise cross sections become available for the large number of transitions required, allowing the magnetic field strengths at which the RWA ceases to be reliable to be established.

%\begin{acknowledgements}
%\end{acknowledgements}

%%%%%%%%%%%%%%%%%%%%%%%%%% APPENDICES %%%%%%%%%%%%%%%%%%%%%%%%%%%
%\appendix
%If you want to present additional material which would interrupt the flow of the main paper,it can be placed in an Appendix which appears after the list of references.
%%%%%%%%%%%%%%%%%%%%%%%%%%%%%%%%%%%%%%%%%%%%%%%%%%

%%%%%%%%%%%%%%%%%%%%%%%%%%%%%%%%%%%%%%%%%%%%%%%%%%%%%%%%%%%%%%%%%
\bibliographystyle{aa}
\bibliography{Hpho}

@ARTICLE{amorim2023,
       author = {{Amorim}, L.~L. and {Kepler}, S.~O. and {K{\"u}lebi}, Baybars and {Jordan}, S. and {Romero}, A.~D.},
        title = "{Catalog of Magnetic White Dwarfs with Hydrogen Dominated Atmospheres}",
      journal = {\apj},
         year = 2023,
        month = feb,
       volume = {944},
       number = {1},
          eid = {56},
        pages = {56},
          doi = {10.3847/1538-4357/acaf6e},
archivePrefix = {arXiv},
       eprint = {2301.08862},
 primaryClass = {astro-ph.SR},
       adsurl = {https://ui.adsabs.harvard.edu/abs/2023ApJ...944...56A}
}

@ARTICLE{alijah1990,
       author = {{Alijah}, Alexander and {Hinze}, Juergen and {Broad}, John T.},
        title = "{Photoionisation of hydrogen in a strong magnetic field}",
      journal = {Journal of Physics B Atomic Molecular Physics},
         year = 1990,
        month = jan,
       volume = {23},
       number = {1},
        pages = {45-60},
          doi = {10.1088/0953-4075/23/1/006},
       adsurl = {https://ui.adsabs.harvard.edu/abs/1990JPhB...23...45A}
}

@ARTICLE{battacharya1985,
       author = {{Bhattacharya}, S.~K. and {Chu}, S. -I.},
        title = "{The resonant photoionisation of hydrogen atom in intense magnetic fields}",
      journal = {Journal of Physics B Atomic Molecular Physics},
         year = 1985,
        month = may,
       volume = {18},
       number = {10},
        pages = {L275-L280},
          doi = {10.1088/0022-3700/18/10/003},
       adsurl = {https://ui.adsabs.harvard.edu/abs/1985JPhB...18L.275B}
}

@BOOK{bethe1957,
       author = {{Bethe}, H.~A. and {Salpeter}, E.~E.},
        title = "{Quantum Mechanics of One- and Two-Electron Atoms}",
         year = 1957,
       adsurl = {https://ui.adsabs.harvard.edu/abs/1957qmot.book.....B}
}

@ARTICLE{burgess1965,
       author = {{Burgess}, A.},
        title = "{Tables of hydrogenic photoionization cross-sections and recombination coefficients}",
      journal = {\memras},
         year = 1965,
        month = jan,
       volume = {69},
        pages = {1},
       adsurl = {https://ui.adsabs.harvard.edu/abs/1965MmRAS..69....1B}
}

@ARTICLE{burkova1976,
       author = {{Burkova}, L.~A. and {Dzyaloshinski{\v{i}}}, I.~E. and
         {Drukarev}, G.~F. and {Monozon}, B.~S.},
        title = "{Hydrogenlike system in crossed electric and magnetic fields}",
      journal = {Soviet Journal of Experimental and Theoretical Physics},
         year = 1976,
        month = aug,
       volume = {44},
        pages = {276},
       adsurl = {https://ui.adsabs.harvard.edu/abs/1976JETP...44..276B}
}

@ARTICLE{canuto1972,
       author = {{Canuto}, V. and {Kelly}, D.~C.},
        title = "{Hydrogen Atom in Intense Magnetic Field}",
      journal = {\apss},
         year = 1972,
        month = aug,
       volume = {17},
       number = {2},
        pages = {277-291},
          doi = {10.1007/BF00642901},
       adsurl = {https://ui.adsabs.harvard.edu/abs/1972Ap&SS..17..277C}
}

@ARTICLE{delande1991,
       author = {{Delande}, D. and {Bommier}, A. and {Gay}, J.~C.},
        title = "{Positive-energy spectrum of the hydrogen atom in a magnetic field}",
      journal = {\prl},
         year = 1991,
        month = jan,
       volume = {66},
       number = {2},
        pages = {141-144},
          doi = {10.1103/PhysRevLett.66.141},
       adsurl = {https://ui.adsabs.harvard.edu/abs/1991PhRvL..66..141D}
}

@ARTICLE{euchner2002,
       author = {{Euchner}, F. and {Jordan}, S. and {Beuermann}, K. and {G{\"a}nsicke}, B.~T. and {Hessman}, F.~V.},
        title = "{Zeeman tomography of magnetic white dwarfs. I. Reconstruction of the field geometry from synthetic spectra}",
      journal = {\aap},
         year = 2002,
        month = aug,
       volume = {390},
        pages = {633-647},
          doi = {10.1051/0004-6361:20020726},
archivePrefix = {arXiv},
       eprint = {astro-ph/0205294},
 primaryClass = {astro-ph},
       adsurl = {https://ui.adsabs.harvard.edu/abs/2002A&A...390..633E}
}

@ARTICLE{garstang1977,
       author = {{Garstang}, R.~H.},
        title = "{REVIEW: Atoms in high magnetic fields (white dwarfs)}",
      journal = {Reports on Progress in Physics},
         year = 1977,
        month = feb,
       volume = {40},
       number = {2},
        pages = {105-154},
          doi = {10.1088/0034-4885/40/2/001},
       adsurl = {https://ui.adsabs.harvard.edu/abs/1977RPPh...40..105G}
}

@ARTICLE{gnedin1974,
       author = {{Gnedin}, Yu. N. and {Pavlov}, G.~G. and {Tsygan}, A.~I.},
        title = "{Photoeffect in strong magnetic fields and x-ray emission from neutron stars}",
      journal = {Soviet Journal of Experimental and Theoretical Physics},
         year = 1974,
        month = aug,
       volume = {39},
        pages = {201},
       adsurl = {https://ui.adsabs.harvard.edu/abs/1974JETP...39..201G}
}

@ARTICLE{grant1958,
       author = {{Grant}, I.~P.},
        title = "{Calculation of Gaunt factors for free-free transitions near positive ions}",
      journal = {\mnras},
         year = 1958,
        month = jan,
       volume = {118},
        pages = {241},
          doi = {10.1093/mnras/118.3.241},
       adsurl = {https://ui.adsabs.harvard.edu/abs/1958MNRAS.118..241G}
}

@ARTICLE{greene1983,
       author = {{Greene}, C.~H.},
        title = "{Atomic photoionization in a strong magnetic field}",
      journal = {\pra},
         year = 1983,
        month = oct,
       volume = {28},
       number = {4},
        pages = {2209-2216},
          doi = {10.1103/PhysRevA.28.2209},
       adsurl = {https://ui.adsabs.harvard.edu/abs/1983PhRvA..28.2209G}
}

@ARTICLE{gorkov1968,
       author = {{Gor'kov}, L.~P. and {Dzyaloshinski{\v{i}}}, I.~E.},
        title = "{Contribution to the Theory of the Mott Exciton in a Strong Magnetic Field}",
      journal = {Soviet Journal of Experimental and Theoretical Physics},
         year = 1968,
        month = feb,
       volume = {26},
        pages = {449},
       adsurl = {https://ui.adsabs.harvard.edu/abs/1968JETP...26..449G}
}

@ARTICLE{guth1929,
       author = {{Guth}, E.},
        title = "{Notiz {\"u}ber den normalen quadratischen Zeemaneffekt}",
      journal = {Zeitschrift fur Physik},
         year = 1929,
        month = may,
       volume = {58},
       number = {5-6},
        pages = {368-372},
          doi = {10.1007/BF01340387},
       adsurl = {https://ui.adsabs.harvard.edu/abs/1929ZPhy...58..368G}
}

@ARTICLE{hatanaka1946,
       author = {{Hatanaka}, T.},
        title = "{Theory of Optical Interaction among He II, OIII and N III Atoms in a Planetary Nebula}",
      journal = {Japanese Journal of Astronomy and Geophysics},
         year = 1946,
        month = jan,
       volume = {21},
        pages = {1},
       adsurl = {https://ui.adsabs.harvard.edu/abs/1946JaJAG..21....1H}
}

@INCOLLECTION{jordan1989,
       author = {{Jordan}, Stefan},
        title = "{Synthetic Spectra of Magnetic White Dwarfs}",
    booktitle = {IAU Colloq. 114: White Dwarfs},
         year = 1989,
       editor = {{Wegner}, Gary},
       volume = {328},
        pages = {333},
          doi = {10.1007/3-540-51031-1_342},
       adsurl = {https://ui.adsabs.harvard.edu/abs/1989LNP...328..333J}
}

@ARTICLE{jordan1992,
       author = {{Jordan}, S.},
        title = "{Models of white dwarfs with high magnetic fields.}",
      journal = {\aap},
         year = 1992,
        month = nov,
       volume = {265},
        pages = {570-576},
       adsurl = {https://ui.adsabs.harvard.edu/abs/1992A&A...265..570J}
}

@ARTICLE{kravchenko1996,
       author = {{Kravchenko}, Yu. P. and {Liberman}, M.~A. and {Johansson}, B.},
        title = "{Exact solution for a hydrogen atom in a magnetic field of arbitrary strength}",
      journal = {\pra},
         year = 1996,
        month = jul,
       volume = {54},
       number = {1},
        pages = {287-305},
          doi = {10.1103/PhysRevA.54.287},
       adsurl = {https://ui.adsabs.harvard.edu/abs/1996PhRvA..54..287K}
}

@INPROCEEDINGS{lamb1972,
       author = {{Lamb}, F.~K. and {Sutherland}, P.~G.},
        title = "{Line Spectra and Continuum Polarization in Magnetic White Dwarfs}",
    booktitle = {Line Formation in the Presence of Magnetic Fields},
         year = 1972,
        month = jan,
        pages = {183},
       adsurl = {https://ui.adsabs.harvard.edu/abs/1972lfpm.conf..183L}
}

@INPROCEEDINGS{lamb1974,
       author = {{Lamb}, F.~K. and {Sutherland}, P.~G.},
        title = "{Continuum Polarization in Magnetic White Dwarfs}",
    booktitle = {Physics of Dense Matter},
         year = 1974,
       editor = {{Hansen}, Carl J.},
       volume = {53},
        month = jan,
        pages = {265},
       adsurl = {https://ui.adsabs.harvard.edu/abs/1974IAUS...53..265L}
}

@ARTICLE{martin1986,
       author = {{Martin}, B. and {Wickramasinghe}, D.~T.},
        title = "{A Test of the Dipole Model for the Rotating Magnetic White Dwarf Feige 7}",
      journal = {\apj},
         year = 1986,
        month = feb,
       volume = {301},
        pages = {177},
          doi = {10.1086/163885},
       adsurl = {https://ui.adsabs.harvard.edu/abs/1986ApJ...301..177M}
}

@ARTICLE{menzel1935,
       author = {{Menzel}, D.~H. and {Pekeris}, C.~L.},
        title = "{Absorption coefficients and hydrogen line intensities}",
      journal = {\mnras},
         year = 1935,
        month = nov,
       volume = {96},
        pages = {77},
          doi = {10.1093/mnras/96.1.77},
       adsurl = {https://ui.adsabs.harvard.edu/abs/1935MNRAS..96...77M}
}

@ARTICLE{meinhardt1999,
       author = {{Meinhardt}, G. and {Schweizer}, W. and {Herold}, H. and {Wunner}, G.},
        title = "{Photoionization of the hydrogen atom in strong magnetic fields of White Dwarfs}",
      journal = {European Physical Journal D},
         year = 1999,
        month = jan,
       volume = {5},
       number = {1},
        pages = {23-31},
          doi = {10.1007/s100530050224},
       adsurl = {https://ui.adsabs.harvard.edu/abs/1999EPJD....5...23M}
}

@ARTICLE{merani1995,
       author = {{Merani}, N. and {Main}, J. and {Wunner}, G.},
        title = "{Balmer and Paschen bound-free opacities for hydrogen in strong white dwarf magnetic fields.}",
      journal = {\aap},
         year = 1995,
        month = jun,
       volume = {298},
        pages = {193},
       adsurl = {https://ui.adsabs.harvard.edu/abs/1995A&A...298..193M}
}

@ARTICLE{mota2007,
       author = {{Mota-Furtado}, F. and {O'Mahony}, P.~F.},
        title = "{R -matrix propagation with adiabatic bases for the photoionization spectra of atoms in magnetic fields}",
      journal = {\pra},
         year = 2007,
        month = nov,
       volume = {76},
       number = {5},
          eid = {053405},
        pages = {053405},
          doi = {10.1103/PhysRevA.76.053405},
archivePrefix = {arXiv},
       eprint = {0709.0275},
 primaryClass = {physics.atom-ph},
       adsurl = {https://ui.adsabs.harvard.edu/abs/2007PhRvA..76e3405M}
}

@ARTICLE{pavlov1980,
       author = {{Pavlov-Verevkin}, V.~B. and {Zhilinskii}, B.~I.},
        title = "{Neutral hydrogen-like system in a magnetic field}",
      journal = {Physics Letters A},
         year = 1980,
        month = aug,
       volume = {78},
       number = {3},
        pages = {244-245},
          doi = {10.1016/0375-9601(80)90082-1},
       adsurl = {https://ui.adsabs.harvard.edu/abs/1980PhLA...78..244P}
}

@ARTICLE{pavlov1993,
       author = {{Pavlov}, G.~G. and {Meszaros}, P.},
        title = "{Finite-Velocity Effects on Atoms in Strong Magnetic Fields and Implications for Neutron Star Atmospheres}",
      journal = {\apj},
         year = 1993,
        month = oct,
       volume = {416},
        pages = {752},
          doi = {10.1086/173274},
       adsurl = {https://ui.adsabs.harvard.edu/abs/1993ApJ...416..752P}
}

@ARTICLE{potekhin1997,
       author = {{Potekhin}, Alexander Y. and {Pavlov}, George G.},
        title = "{Photoionization of Hydrogen in Atmospheres of Magnetic Neutron Stars}",
      journal = {\apj},
         year = 1997,
        month = jul,
       volume = {483},
       number = {1},
        pages = {414-425},
          doi = {10.1086/304250},
archivePrefix = {arXiv},
       eprint = {astro-ph/9702004},
 primaryClass = {astro-ph},
       adsurl = {https://ui.adsabs.harvard.edu/abs/1997ApJ...483..414P}
}

@ARTICLE{putney1995,
       author = {{Putney}, Angela and {Jordan}, Stefan},
        title = "{Off-centered Dipole Models for Three Isolated Magnetic White Dwarfs}",
      journal = {\apj},
         year = 1995,
        month = aug,
       volume = {449},
        pages = {863},
          doi = {10.1086/176103},
       adsurl = {https://ui.adsabs.harvard.edu/abs/1995ApJ...449..863P}
}

@ARTICLE{rohrmann2025,
       author = {{Rohrmann}, Ren{\'e} D.},
        title = "{Strong signature of right-handed circularly polarized photoionization close to the cyclotron line in the atmosphere of magnetic white dwarfs}",
      journal = {\aap},
         year = 2025,
        month = jan,
       volume = {693},
          eid = {L5},
        pages = {L5},
          doi = {10.1051/0004-6361/202452569},
archivePrefix = {arXiv},
       eprint = {2412.06627},
 primaryClass = {astro-ph.SR},
       adsurl = {https://ui.adsabs.harvard.edu/abs/2025A&A...693L...5R}
}

@ARTICLE{rozsnyai1988,
       author = {{Rozsnyai}, Balazs F. and {Jacobs}, Verne L.},
        title = "{Photorecombination Rates of Hydrogenic and Nonhydrogenic States}",
      journal = {\apj},
         year = 1988,
        month = apr,
       volume = {327},
        pages = {485},
          doi = {10.1086/166211},
       adsurl = {https://ui.adsabs.harvard.edu/abs/1988ApJ...327..485R}
}

@ARTICLE{schiff1939,
       author = {{Schiff}, L.~I. and {Snyder}, H.},
        title = "{Theory of the Quadratic Zeeman Effect}",
      journal = {Physical Review},
         year = 1939,
        month = jan,
       volume = {55},
       number = {1},
        pages = {59-63},
          doi = {10.1103/PhysRev.55.59},
       adsurl = {https://ui.adsabs.harvard.edu/abs/1939PhRv...55...59S}
}

@ARTICLE{schimeczek2014,
       author = {{Schimeczek}, C. and {Wunner}, G.},
        title = "{Atomic Data for the Spectral Analysis of Magnetic DA White Dwarfs in the SDSS}",
      journal = {\apjs},
         year = 2014,
        month = jun,
       volume = {212},
       number = {2},
          eid = {26},
        pages = {26},
          doi = {10.1088/0067-0049/212/2/26},
       adsurl = {https://ui.adsabs.harvard.edu/abs/2014ApJS..212...26S}
}

@ARTICLE{schmidt1981,
       author = {{Schmidt}, W. and {Herold}, H. and {Ruder}, H. and {Wunner}, G.},
        title = "{The Photoionisation of the Hydrogen Atom in Strong Magnetic Fields}",
      journal = {\aap},
         year = 1981,
        month = jan,
       volume = {94},
        pages = {194},
       adsurl = {https://ui.adsabs.harvard.edu/abs/1981A&A....94..194S}
}

@ARTICLE{simola1978,
       author = {{Simola}, J. and {Virtamo}, J.},
        title = "{Energy levels of hydrogen atoms in a strong magnetic field}",
      journal = {Journal of Physics B Atomic Molecular Physics},
         year = 1978,
        month = oct,
       volume = {11},
       number = {19},
        pages = {3309-3322},
          doi = {10.1088/0022-3700/11/19/008},
       adsurl = {https://ui.adsabs.harvard.edu/abs/1978JPhB...11.3309S}
}

@ARTICLE{stobbe1930,
       author = {{Stobbe}, M.},
        title = "{Zur Quantenmechanik photoelektrischer Prozesse}",
      journal = {Annalen der Physik},
         year = 1930,
        month = jan,
       volume = {399},
       number = {6},
        pages = {661-715},
          doi = {10.1002/andp.19303990604},
       adsurl = {https://ui.adsabs.harvard.edu/abs/1930AnP...399..661S}
}

@BOOK{vanvleck1932,
       author = {{Van Vleck}, J.~H.},
        title = "{Theory of Electric and Magnetic Susceptibilities}",
         year = 1932 
}

@ARTICLE{vera2020,
       author = {{Vera-Rueda}, Mat{\'\i}as and {Rohrmann}, Ren{\'e} D.},
        title = "{Hydrogen ionization equilibrium in magnetic fields}",
      journal = {\aap},
         year = 2020,
        month = mar,
       volume = {635},
          eid = {A180},
        pages = {A180},
          doi = {10.1051/0004-6361/201937413},
       adsurl = {https://ui.adsabs.harvard.edu/abs/2020A&A...635A.180V}
}

@ARTICLE{vera2024,
       author = {{Vera-Rueda}, Mat{\'\i}as and {Rohrmann}, Ren{\'e} D.},
        title = "{A numerical code for the analysis of magnetic white-dwarf spectra that includes field effects on the chemical equilibrium}",
      journal = {\aap},
         year = 2024,
        month = jul,
       volume = {687},
          eid = {A141},
        pages = {A141},
          doi = {10.1051/0004-6361/202449627},
       adsurl = {https://ui.adsabs.harvard.edu/abs/2024A&A...687A.141V}
}

@ARTICLE{vincke1988,
       author = {{Vincke}, M. and {Baye}, D.},
        title = "{Centre-of-mass effects on the hydrogen atom in a magnetic field}",
      journal = {Journal of Physics B Atomic Molecular Physics},
         year = 1988,
        month = jul,
       volume = {21},
       number = {13},
        pages = {2407-2424},
          doi = {10.1088/0953-4075/21/13/009},
       adsurl = {https://ui.adsabs.harvard.edu/abs/1988JPhB...21.2407V}
}

@ARTICLE{wang1991,
       author = {{Wang}, Qiaoling and {Greene}, Chris H.},
        title = "{R-matrix calculation of atomic hydrogen photoionization in a strong magnetic field}",
      journal = {\pra},
         year = 1991,
        month = dec,
       volume = {44},
       number = {11},
        pages = {7448-7458},
          doi = {10.1103/PhysRevA.44.7448},
       adsurl = {https://ui.adsabs.harvard.edu/abs/1991PhRvA..44.7448W}
}

@INPROCEEDINGS{wickramasinghe1995,
       author = {{Wickramasinghe}, D.~T.},
        title = "{Opacities in Strong Magnetic Fields and the Atmospheres of Magnetic White Dwarfs}",
    booktitle = {Astrophysical Applications of Powerful New Databases},
         year = 1995,
       editor = {{Adelman}, Saul J. and {Wiese}, W.~L.},
       series = {Astronomical Society of the Pacific Conference Series},
       volume = {78},
        month = jan,
        pages = {319},
       adsurl = {https://ui.adsabs.harvard.edu/abs/1995ASPC...78..319W}
}

@ARTICLE{wickramasinghe2000,
       author = {{Wickramasinghe}, D.~T. and {Ferrario}, Lilia},
        title = "{Magnetism in Isolated and Binary White Dwarfs}",
      journal = {\pasp},
         year = 2000,
        month = jul,
       volume = {112},
       number = {773},
        pages = {873-924},
          doi = {10.1086/316593},
       adsurl = {https://ui.adsabs.harvard.edu/abs/2000PASP..112..873W}
}

@ARTICLE{zhao2007,
       author = {{Zhao}, L.~B. and {Stancil}, P.~C.},
        title = "{Hydrogen Photoionization Cross Sections for Strong-Field Magnetic White Dwarfs}",
      journal = {\apj},
         year = 2007,
        month = oct,
       volume = {667},
       number = {2},
        pages = {1119-1125},
          doi = {10.1086/520948},
       adsurl = {https://ui.adsabs.harvard.edu/abs/2007ApJ...667.1119Z}
}

@ARTICLE{zhao2021,
       author = {{Zhao}, L.~B.},
        title = "{Lyman and Balmer Continuum Spectra for Hydrogen Atoms in Strong White Dwarf Magnetic Fields}",
      journal = {\apjs},
         year = 2021,
        month = jun,
       volume = {254},
       number = {2},
          eid = {21},
        pages = {21},
          doi = {10.3847/1538-4365/abf6d0},
       adsurl = {https://ui.adsabs.harvard.edu/abs/2021ApJS..254...21Z}
}
\end{document}